\documentclass{article}

\usepackage{arxiv}

\usepackage{amsmath}
\usepackage{graphicx}
\usepackage{epstopdf}
\usepackage[utf8]{inputenc} 
\usepackage[T1]{fontenc}    
\usepackage{hyperref}       
\usepackage{url}            
\usepackage{booktabs}       
\usepackage{amsfonts}       
\usepackage{nicefrac}       
\usepackage{microtype}      
\usepackage{lipsum}
\usepackage{algorithm}
\usepackage{algorithmic}
\usepackage{booktabs}
\usepackage{lscape}
\usepackage{caption}
\usepackage{subcaption}
\usepackage{multirow}
\newtheorem{thm}{Theorem}[section]

\numberwithin{equation}{section} \allowdisplaybreaks

\title{Mathematical Modelling of Streamwise Velocity Profile in Open Channels using Tsallis Entropy}

\author{
 Manotosh Kumbhakar \\
  School of Basic Sciences\\ 
  Indian Institute of Technology Mandi\\
  Mandi 175005, India\\
  \texttt{manotosh.kumbhakar@gmail.com} \\
    \And
  Rajendra K. Ray \\
  School of Basic Sciences\\ 
  Indian Institute of Technology Mandi\\
  Mandi 175005, India\\
  \texttt{rajendra@iitmandi.ac.in} \\
    \And
   Suvra Kanti Chakraborty \\
   Department of Mathematics\\ 
   Sir Gurudas Mahavidyalaya\\ 
   Kolkata 700067, India\\
  \texttt{suvrakanti89@gmail.com} \\ 
  \And 
   Koeli Ghoshal \\
   Department of Mathematics\\ 
   Indian Institute of Technology Kharagpur\\ 
   Kharagpur 721302, India\\
   \texttt{koeli@maths.iitkgp.ac.in} \\ 
    \And 
   Vijay P. Singh \\
   Department of Biological and Agricultural Engineering\\
   Texas A\&M University\\
   College Station TX 77843-2117, USA\\
  \texttt{vsingh@tamu.edu} \\ 
}

\begin{document}
\maketitle

\begin{abstract}
This study derived the vertical distribution of streamwise velocity in wide open channels by maximizing Tsallis entropy, in accordance with the maximum entropy principle, subject to the total probability rule and the conservation of mass, momentum, and energy. Entropy maximizing leads to a highly nonlinear differential equation for velocity which was transformed into a relatively weaker nonlinear equation and then solved analytically using a non-perturbation approach that yielded a series solution. The convergence of the series solution was proved using both theoretical and numerical procedures. For the assessment of velocity profile, the Lagrange multipliers and the entropy index were obtained by solving a system of nonlinear equations by Gauss-Newton method after approximating the constraint integrals using Gauss-Legendre quadrature rule. The derived velocity profile was validated for some selected sets of experimental and field data and also compared with the existing velocity profile based on Tsallis entropy. The incorporation of the above constraints and the effect of entropy index were found to improve the velocity profile for experimental as well as field data. The methodology reported in this study can also be employed for addressing other open channel flow problems, such as sediment concentration and shear stress distribution.
\end{abstract}

\keywords{Maximum entropy principle \and Tsallis distribution \and Pad{\'e} approximation \and Open channel flow \and Analytical solution.}

\section{Introduction}
\label{intro}
Investigation into the velocity distribution in open channel turbulent flow leads to a variety of applications in the field of sediment transport and hydrodynamics \cite{chow1959open,vanoni2006sedimentation}. An open channel is a conduit having free water surface in contact with the atmosphere, such as rivers, streams, canals, ditches, etc. Transport phenomena in rivers/canals require velocity/sediment concentration distribution often investigated empirically through laboratory and field experiments as well as theoretically.
\par
The vertical velocity profile was proposed by Prandtl \cite{prandtl19257} known as the classical log-law which is valid mostly in the inner region ($y/D<0.2$) and shows a deviation from the experimental data in the outer region ($y/D>0.2$) of the channel \cite{nezu2005open}, where $D$ is the flow depth and $y$ is the vertical distance from the channel bed. The discrepancy of the log-law in the outer region was addressed by Coles \cite{coles1956law} through the wake function and later both log and wake laws were modified \cite{guo2003modified,yang2005investigation}. Apart from log-law and log wake law, power law \cite{afzal2005scaling} is also widely used to describe the velocity distribution along a vertical. However, these laws are all valid for wide-open channels, and for narrow open channels where the maximum velocity occurs some distance below the free surface, both log law and power law fail to describe the velocity distribution. Several researchers \cite{yang2004velocity,guo2003modified,kundu2012analytical} modified the log-law to describe the velocity distribution in narrow open channels. 
\par
The aforementioned studies employ deterministic methods that are based on the solution of RANS equation, and any numerical model encounters difficulty in computation and the accuracy of solution is often questionable. Also, that in the deterministic methods, the time-averaged velocity is considered and the uncertainty in flow is ignored. To overcome these limitations, a probabilistic approach using information theory was applied first by Chiu \cite{chiu1987entropy} to the study the distributions of velocity, shear stress, and sediment concentration. Chiu's approach employed the Shannon entropy theory together with the principle of maximum entropy \cite{shannon1948mathematical,jaynesa1957information,jaynesb1957information}. The maximum entropy approach has been extensively studied in different branches of science and engineering \cite{kapur1989maximum}. The concept of Chiu was further employed to establish the relationship between mean and maximum velocities through an entropy parameter for determining discharge in streams and rivers \cite{chiu1995maximum}, to determine the cross-section of a straight threshold channel \cite{cao1997entropy}, to estimate the depth-averaged sediment concentration based on the vertical concentration distribution \cite{chiu2000mathematical}, to derive the transverse distribution of boundary shear stress in circular open channels \cite{sterling2002attempt}, to calculate the hillslope sediment production \cite{araujo2007entropy}, to link Chiu's entropy parameter and the geometric and hydraulic characteristics of river cross-sections \cite{moramarco2010formulation}, to estimate river discharge at the Danshui River, the largest estuarine system in Taiwan \cite{bechle2014entropy}, to find a relationship between Chiu's entropy parameter and the relative submergence in open-channel flows with large-scale roughness \cite{greco2014entropy}, to formulate a model on velocity-dip-position in a flow cross-section \cite{kundu2017prediction}, and so on \cite{singh2014entropy}.
\par
The generalization of Shannon entropy was formulated by Tsallis \cite{tsallis1988possible}, which is a non-extensive entropy and contains an additional parameter (often called entropy index). Tsallis entropy has several important properties like it takes on maximum value in the case of equiprobability and is pseudo-additive for independent subsystems, and these properties make Tsallis entropy important in the field of science and engineering (details is available at \url{http://tsallis.cat.cbpf.br/biblio.htm}). In hydraulics, following the work of Chiu \cite{chiu1987entropy}, the concept of Tsallis entropy was explored for studying 1D and 2D velocity distributions in open channels by \cite{singh2011entropy} and \cite{luo2010entropy}, respectively. They found that the velocity profile obtained from Tsallis entropy captured the near-bed data better than that from Shannon entropy. The idea was further extended by Cui and Singh \cite{cui2013one,cui2012two} with some modification to the hypothesized cumulative distribution function (CDF) in the space domain. Besides studies on velocity, Tsallis entropy was successfully applied to other problems of hydraulics and hydrology also \cite{koutsoyiannisa2005uncertainty,koutsoyiannisb2005uncertainty,keylock2005describing,bonakdari2015comparison,singh2016introduction,gholami2019method}. However, studies on open channel flow velocity using Tsallis entropy considered, for mathematical convenience, the constraints based only on the total probability rule and mass conservation, and also concluded a fixed value of the entropy index based on a data-fitting procedure without any physical justification. It can be concluded from the literature that in general the Tsallis entropy index is considered to be an adjustable parameter which leads to a proper resulting distribution while validating with experimental observations \cite{tsallis1999nonextensive}, except for a few cases such as \cite{lyra1998nonextensivity} and \cite{conroy2015determining}. Therefore, the primary objective of the present study is to derive a streamwise velocity profile in open channels using Tsallis entropy by incorporating the constraints based on the total probability rule and the conservation of mass, momentum, and energy. In addition, the study includes the effect of entropy index for modelling velocity in open channels rather than treating it as a fixed value. Since most of the open channel flow problems do not contain a small parameter in the governing equation, the differential equation is solved using a non-perturbation approach. Further, the effect of additional constraints and the entropy index is investigated through sets of experimental and field data available in the literature and by comparing the derived velocity profile with the existing velocity profile. 

\section{Mathematical Formulation}
\label{sec:main}
\subsection{Probability Distribution of Velocity}
Let us consider an open channel with a flow of depth $D$ and width $B$ where the time-averaged normalized streamwise velocity $\hat{u}$ is assumed to be a random variable having the PDF $f(\hat{u})$. Then, the Tsallis entropy (\cite{tsallis1988possible}) of $\hat{u}$ can be written in a continuous (differential) form as
\begin{equation}\label{eq1}
    {H_{{q_0}}}\left[ {f\left( {\hat u} \right)} \right] = \frac{1}{{{q_0} - 1}}\mathop \int \limits_{\hat u \in \Theta } f\left( {\hat u} \right)\left[ {1 - {{\left( {f\left( {\hat u} \right)} \right)}^{{q_0} - 1}}} \right]d\hat u
\end{equation}
where $\hat u = u/{u_{max}}$, ${u_{max}}$ being the maximum velocity in a given flow cross-section, $\Theta  = \left[ {0,1} \right]$ is the domain of $\hat u$, ${q_0}$ is the real parameter known as Tsallis entropy index. In the limit as ${q_0} \to 1$ in Eq. (\ref{eq1}), the Shannon entropy is recovered. Tsallis entropy attains its maximum when the PDF $f\left( {\hat u} \right)$ is uniform for any value of ${q_0}$. Moreover, the function ${H_{{q_0}}}$ is concave for ${q_0} > 0$ and convex for ${q_0} < 0$ (\cite{plastino1999tsallis}). 
\par
The objective is to determine the velocity distribution by applying the principle of maximum entropy to Tsallis entropy Eq. (\ref{eq1}), subject to the specified constraints. The constraints can be prescribed in accordance with the conservation of mass, momentum, and energy which can be computed from observations. The PDF $f(\hat{u})$ satisfies the total probability rule, i.e., 
\begin{equation}\label{eq2}
  \mathop \int \limits_{\hat u \in \Theta } f\left( {\hat u} \right)d\hat u = 1  
\end{equation}
Now the constraints based on conservation of mass, momentum, and energy can be defined, respectively, as
\begin{equation}\label{eq3}
    {m_1} = \mathop \int \limits_{\hat u \in \Theta } \hat uf\left( {\hat u} \right)d\hat u = \overline {\hat u} 
\end{equation}
\begin{equation}\label{eq4}
    {m_2} = \mathop \int \limits_{\hat u \in \Theta } {\hat u^2}f\left( {\hat u} \right)d\hat u = \overline {{{\hat u}^2}}  = \beta {\overline {\hat u} ^2}
\end{equation}
\begin{equation}\label{eq5}
    {m_3} = \mathop \int \limits_{\hat u \in \Theta } {\hat u^3}f\left( {\hat u} \right)d\hat u = \overline {{{\hat u}^3}}  = \alpha {\overline {\hat u} ^3}
\end{equation}
where $\overline {\hat u}  = \bar u/{u_{max}}$ is the dimensionless mean velocity, $\beta$ is the momentum distribution coefficient, and $\alpha$ is the energy distribution coefficient. It has been shown that the first-, second-, and third-order moments given by Eqs. (\ref{eq3}), (\ref{eq4}), and (\ref{eq5}) represent the hydrodynamic transport of mass, momentum, and energy, respectively (\cite{chiu1989velocity,kumbhakar2019application}). The higher-order moments can indeed be defined similarly; however, for velocity modeling Eqs. (\ref{eq2})-(\ref{eq5}) represent a complete set of constraints as they are based on the total probability rule and the basic conservation laws. 
\par
Following Jaynes’ principle of maximum entropy (\cite{jaynesa1957information,jaynesb1957information}), the velocity distribution can be obtained by maximizing the Tsallis entropy function Eq. (\ref{eq1}) subject to the constraints Eqs. (\ref{eq2})-(\ref{eq5}). To that end the Lagrangian function is constructed as
\begin{align}\label{eq6}
    L\left( {f,\boldsymbol{\lambda} } \right) = \frac{1}{{{q_0} - 1}}\mathop \int \nolimits_{\hat u \in \Theta } & f\left( {\hat u} \right)\left[ {1 - {{\left( {f\left( {\hat u} \right)} \right)}^{{q_0} - 1}}} \right]d\hat u + {\lambda _0}\left( {\mathop \int \nolimits_{\hat u \in \Theta } f\left( {\hat u} \right)d\hat u - 1} \right) \nonumber \\&+  \mathop \sum \limits_{i = 1}^{{N_0}} {\lambda _i}\left( {\mathop \int \nolimits_{\hat u \in \Theta } {{\hat u}^i}f\left( {\hat u} \right)d\hat u - {m_i}} \right)
\end{align}
where ${N_0} =3$, the number of constraints; and ${\lambda _i}$, $i = 0,1, \ldots ,{N_0}$ are the Lagrange multipliers. Application of the Euler-Lagrange equation $\frac{{\partial L}}{{\partial f}} - \frac{d}{{d\hat u}}\left( {\frac{{\partial L}}{{\partial f'}}} \right) = 0$ to Eq. (\ref{eq6}) produces the PDF of velocity as
\begin{equation}\label{eq7}
    f\left( {\hat u} \right) = {\left( {\frac{{{q_0} - 1}}{{{q_0}}}\left[ {\frac{1}{{{q_0} - 1}} + \mathop \sum \limits_{i = 0}^{{N_0}} {\lambda _i}{{\hat u}^i}} \right]} \right)^{\frac{1}{{{q_0} - 1}}}}
\end{equation}
The cumulative distribution function (CDF) of velocity can be obtained from Eq. (\ref{eq7}) as
\begin{equation}\label{eq8}
    F\left( {\hat u} \right) = Prob\left( {\hat U \le \hat u} \right) = \mathop \int \limits_0^{\hat u} f\left( {\hat u} \right)d\hat u = \mathop \int \limits_0^{\hat u} {\left( {\frac{{{q_0} - 1}}{{{q_0}}}\left[ {\frac{1}{{{q_0} - 1}} + \mathop \sum \limits_{i = 0}^{{N_0}} {\lambda _i}{{\hat u}^i}} \right]} \right)^{\frac{1}{{{q_0} - 1}}}}d\hat u
\end{equation}

\subsection{Connection with the Space Domain}
The aim is to find the spatial distribution of streamwise velocity in open channels. The nature of velocity distribution in open channels exhibits different characteristics based on channel-type. The ratio of channel width $B$ to flow depth $D$, known as aspect ratio, divides open channels into two categories, namely, wide and narrow channels. When $B/D <5$ the channel is narrow, for $B/D>10$ it is wide, and for other cases, it depends on the nature of the surface roughness. In narrow open channels, the maximum velocity occurs below the water surface due to the presence of strong secondary currents that arise by the effect of the sidewall of the channel (\cite{chow1959open}). As the primary objective of the present work is to check the effects of constraints and the entropy index on the velocity profile, we restrict our analysis to the case of wide channels where the velocity varies in the vertical direction only, and the maximum velocity appears at the water surface. Now, the connection between the probability and the space domain is made by assuming that there exists a relation of the type
\begin{equation}\label{eq9}
    u = g\left( y \right),\;\;g\;{\rm{being\;an\;arbitrary\;function\;}}
\end{equation}
It can be assumed that all the values of $y$ between 0 to $D$ are equally likely so that PDF of $y$ follows a uniform distribution
\begin{equation}\label{eq10}
    {f_1}\left( y \right) = \left\{ {\begin{array}{*{20}{c}}{\frac{1}{D}\;\;\;\;{\rm{for}}\;\;\;0 \le y \le D}\\{0\;\;\;\;\;\;\;\;\;\;\;\;\;\;\;{\rm{otherwise}}}\end{array}} \right.
\end{equation}
Using Eq. (\ref{eq9}), the distribution of $u$ can be written as
\begin{equation}\label{eq11}
    f\left( u \right) = {f_1}\left( y \right)\left| {\frac{{dy}}{{du}}} \right| = \frac{1}{D}\frac{{dy}}{{du}}
\end{equation}
Using the definition, the PDF of $\hat{u}$ can be obtained from Eq. (\ref{eq11}) as follows
\begin{equation}\label{eq12}
    f\left( {\hat u} \right) = {u_{max}}f\left( u \right) = \frac{{d\hat y}}{{d\hat u}}
\end{equation}
Further, the CDF can be obtained as
\begin{equation}\label{eq13}
    F\left( {\hat u} \right) = Prob\left( {\hat U \le \hat u} \right) = \mathop \int \limits_0^{\hat u} f\left( {\hat u} \right)d\hat u = \mathop \int \limits_{{g^{ - 1}}\left( 0 \right)}^{{g^{ - 1}}\left( {\hat u} \right)} f\left( {g\left( y \right)} \right)\frac{{d\hat u}}{{dy}}dy = \mathop \int \limits_0^y {f_1}\left( y \right)dy = \frac{y}{D}
\end{equation}
Physically Eq. (\ref{eq13}) represents the fraction of total cross-section area in which the velocity is less than or equal to $\hat{u}$. It may be noted that this concept can also be extended  to the case of two-dimensional distribution of velocity by hypothesizing a generalized coordinate system (\cite{chiu1989velocity,kumbhakar2019application}). 

\subsection{Governing Differential Equation for Velocity }
Equating Eqs. (\ref{eq7}) and (\ref{eq12}), the streamwise velocity profile is governed by the following differential equation:
\begin{equation}\label{eq14}
    \frac{{d\hat u}}{{d\hat y}} = {\left( {\frac{{{q_0} - 1}}{{{q_0}}}\left[ {\frac{1}{{{q_0} - 1}} + \mathop \sum \limits_{i = 0}^{{N_0}} {\lambda _i}{{\hat u}^i}} \right]} \right)^{\frac{1}{{1 - {q_0}}}}}\;\;{\rm{subject\;to}}\;\;\hat u\left( {y = 0} \right) = 0
\end{equation}
Eq. (\ref{eq14}) is a highly non-linear ordinary differential equation, and depends on the entropy index and the Lagrange multipliers. An analytical solution of the equation may be a challenging task due to the strong nonlinear term on the right side. The determination of entropy index and Lagrange multipliers is essential for assessing the velocity profile.
\subsection{Determination of Entropy Index and Lagrange Multipliers}
For determining the Lagrange multipliers, one can substitute the PDF given by Eq. (\ref{eq7}) in the constraint Eqs. (\ref{eq2})-(\ref{eq5}). Apart from the multipliers, the entropy index is an additional parameter when modelling velocity using Tsallis entropy. Previous studies included only the mass conservation constraint, and considering test cases for the entropy index $q_0$, they concluded its best value to be 3/4 and 2 for 1D and 2D velocity distributions, respectively (\cite{singh2011entropy,luo2010entropy}). Later, Cui and Singh (\cite{cui2012two,cui2013one}) extended their work by introducing some empirical parameters in the hypothesized CDF and determined the best choice for $q_0$ to be 3 for both 1D and 2D velocity distributions. However, their approaches are only heuristic, based on data, and do not provide any physical justification for the entropy index. Here we aim to find the physical justification for the entropy index and hence we incorporate three constraints to formulate the velocity distribution model, i.e., $N_0=2$ and use the energy constraint to close the system. This way the constraint equations lead to a system of four equations with four unknowns $\lambda_i$'s for $i=0,1,2$ and $q_0$, and a complete set of constraints is included for analyzing the velocity profile.  
\par
Putting $N_0=2$ in Eq. (\ref{eq7}), $f(\hat{u})$ can be substituted in the constraints to determine the parameters. However, the integrands may not be performed analytically. To that end, first the Gauss-Legendre quadrature rule is applied to approximate the integrals, and then a system of four equations with four unknowns $\lambda_i$'s and $q_0$ is obtained. The Gaussian quadrature rule approximates the definite integral of a function as a weighted sum of functional values at some specified points within the domain of the integration. A general $N$-point Gaussian-Legendre quadrature rule for the function $\Psi$ over the domain $[-1,1]$ can be written as
\begin{equation}\label{eq15}
    \mathop \int \limits_{ - 1}^1 \Psi \left( x \right)dx = \mathop \sum \limits_{K = 1}^N {w_{N,K}}\Psi \left( {{x_{N,K}}} \right) + {E_N}\left( \Psi  \right)
\end{equation}
where ${x_{N,K}}$ and ${w_{N,K}}$ are the nodes and weights, respectively, for $K = 1,2, \ldots ,N$. The first term on the right side of Eq. (\ref{eq15}) represents the numerical approximation of the integral and the second term is the error. Eq. (\ref{eq15}) is exact for polynomial functions of degree less than or equal to $2N - 1$. To apply the rule over an arbitrary interval $\left[ {{\eta _1},{\eta _2}} \right]$, one can simply use the change of the variable $t = \frac{{{\eta _1} + {\eta _2}}}{2} + \frac{{{\eta _2} - {\eta _1}}}{2}x$ so that the Gauss-Legendre quadrature becomes
\begin{equation}\label{eq16}
    \resizebox{1\hsize}{!}{$\mathop \int \limits_{{\eta _1}}^{{\eta _2}} \Psi \left( t \right)dt = \frac{{{\eta _2} - {\eta _1}}}{2}\mathop \int \limits_{ - 1}^1 \Psi \left( {\frac{{{\eta _1} + {\eta _2}}}{2} + \frac{{{\eta _2} - {\eta _1}}}{2}x{\rm{\;}}} \right)dx \approx \mathop \sum \limits_{k = 1}^N {w_{N,K}}\Psi \left( {\frac{{{\eta _1} + {\eta _2}}}{2} + \frac{{{\eta _2} - {\eta _1}}}{2}{x_{N,K}}} \right)$}
\end{equation}
The Gauss-Legendre quadrature formula is a powerful technique for approximating the definite integrals. There are many algorithms available to determine the nodes and weights of the formula. Here we use the MATLAB script available at \url{https://in.mathworks.com/matlabcentral/fileexchange}. This script uses the Legendre-Gauss Vandermonde Matrix, and calculates the nodes and weights by computing the zeros of the Legendre polynomial using recurrence relation along with the Newton-Raphson method. Now, applying the aforementioned technique to the constraints Eqs. (\ref{eq2})-(\ref{eq5}) after substituting the PDF given by Eq. (\ref{eq7}) with $N_0=2$,  one can arrive at the following system of nonlinear equations
\begin{equation}\label{eq17}
    {\varphi _1}\left( {{\lambda _0},{\lambda _1},{\lambda _2},{q_0}} \right) \equiv \;{G_1}\left( {{\lambda _0},{\lambda _1},{\lambda _2},{q_0}} \right) - 1 = 0
\end{equation}
\begin{equation}\label{eq18}
   {\varphi _2}\left( {{\lambda _0},{\lambda _1},{\lambda _2},{q_0}} \right) \equiv {G_2}\left( {{\lambda _0},{\lambda _1},{\lambda _2},{q_0}} \right) - \overline {\hat u}  = 0 
\end{equation}
\begin{equation}\label{eq19}
   {\varphi _3}\left( {{\lambda _0},{\lambda _1},{\lambda _2},{q_0}} \right) \equiv {G_3}\left( {{\lambda _0},{\lambda _1},{\lambda _2},{q_0}} \right) - \beta {\overline {\hat u} ^2} = 0 
\end{equation}
\begin{equation}\label{eq20}
    {\varphi _4}\left( {{\lambda _0},{\lambda _1},{\lambda _2},{q_0}} \right) \equiv {G_4}\left( {{\lambda _0},{\lambda _1},{\lambda _2},{q_0}} \right) - \alpha {\overline {\hat u} ^3} = 0
\end{equation}
where ${G_i}$'s for $i = 1,2,3,4$ are the Gauss-Legendre quadrature-based approximations for the integrals of Eqs. (\ref{eq2})-(\ref{eq5}), respectively. For solving the nonlinear system, Eqs. (\ref{eq17})-(\ref{eq20}) can be represented in vector form as  $\varphi \left( \zeta  \right) = 0$ where $\varphi \left( \zeta  \right) = {\left[ {{\varphi _1}\left( \zeta  \right),\;{\varphi _2}\left( \zeta  \right),{\varphi _3}\left( \zeta  \right),{\varphi _4}\left( \zeta  \right)} \right]^{'}}$ and  $\zeta  = \left[ {{\lambda _0},{\lambda _1},{\lambda _2},{q_0}} \right]$. Here each ${\varphi _i}\left( \zeta  \right)$ for $i = 1,2,3,4$ is smooth. The solution methodology is now described below. 
\par
Newton's method is used for handling the situation in the above framework (\cite{kelley2003solving}). Since \(\varphi :\;{\mathbb{R}^4} \to {\mathbb{R}^4}\) is continuously differentiable, for two vectors \(\zeta \) and \(\zeta  + p\) in the domain, we have 
\begin{equation}\label{eq21}
    \varphi \left( {\zeta  + p} \right) = \varphi \left( \zeta  \right) + \mathop \int \limits_0^1 J\left( {\zeta  + tp} \right)p\;dt
\end{equation}
where $J\left( \zeta  \right) = \varphi '\left( \zeta  \right)$ is the Jacobian matrix. A linear model may be developed by approximating the second term on the right-hand side of Eq. (\ref{eq21}) as $J\left( \zeta  \right)p$, and then can be written in an iterative form as
\begin{equation}\label{eq22}
    {M_k}\left( p \right) = \;\varphi \left( {{\zeta _k}} \right) + J\left( {{\zeta _k}} \right)p
\end{equation}
where ${\zeta _k}$ is the approximate solution of the system at the ${k^{th}}$ iteration. Newton's method in its pure form chooses the step ${p_k}$ such that ${M_k}\left( {{p_k}} \right) = 0$, which eventually results in ${p_k} =  - J{\left( {{\zeta _k}} \right)^{ - 1}}\;\varphi \left( {{\zeta _k}} \right)$. The iterative process is updated as ${\alpha _{k + 1}} = {\alpha _k} + {p_k}$, and then continued until $\left| {\left| {\;\varphi \left( {{\zeta _k}} \right)\;} \right|} \right|$ is smaller than some desired tolerance limit. However, in Newton's method, the computation of $J\left( \zeta  \right)$ may be expensive or sometimes it is difficult to obtain. Also, the method may behave erratically if the starting point is far from the solution point. Besides, if $J\left( {{\zeta _k}} \right)$ is singular for some $k$, the Newton-step ${p_k}$ may not even be defined. To avoid the computation of $J\left( \zeta  \right)$, one may adopt the logic of the quasi-Newton method that approximates $J\left( \zeta  \right)$ at each iteration and mimics the behavior of the true Jacobian matrix (\cite{martinez2000practical}). Let ${B_k}$ be the approximate matrix of $J\left( \zeta  \right)$ at the ${k^{th}}$ iteration, which can be used to form a linear model similar to that of Newton's method as
\begin{equation}\label{eq23}
    {M_k}\left( p \right) = \;\varphi \left( {{\zeta _k}} \right) + {B_k}p
\end{equation}
When ${B_k}$ is nonsingular, one obtains ${p_k} =  - B_k^{ - 1}\varphi \left( {{\zeta _k}} \right)$ after setting Eq. (\ref{eq23}) to zero. Then, at each step, the approximate matrix is modified using Broyden's formula as \begin{equation}\label{eq24}
    {B_{k + 1}} = \;{B_k} + \frac{{\left( {{y_k} - {B_k}{s_k}} \right)s_k^{'}}}{{s_k^{'}{s_k}}}
\end{equation}
where  ${s_k} = {\zeta _{k + 1}} - {\zeta _k},\;\)and \({y_k} = \varphi \left( {{\zeta _{k + 1}}} \right) - \varphi \left( {{\zeta _k}} \right)$. But it is shown (\cite{nocedal2006numerical}) that to ensure the convergence of the quasi-Newton method, an assumption that the initial approximate ${B_0}$ must be close to the Jacobian matrix at the solution ${\zeta ^*}$, is taken.  However, some implementation of this Broyden's quasi-Newton method proposes $B_{0} = J(\zeta_{0})$.
\par
Neither Newton's method nor Broyden's method with unit step length can guarantee to converge to the solution unless the starting point is in the vicinity of the solution. Hence, modification of these methods was required for their practical use, which paves the way of associating a line search technique to the existing concept of Newton-like methods. The line search method searches for an improving point in Newton or quasi-Newton iteration. The most widely used merit function related to the line search methods, which determines the acceptance of the new iteration in the case of solving a system of nonlinear equation, is the sum of squares, i.e., $f\left( \zeta  \right) = \frac{1}{2}{\left| {\left| {\varphi \left( \zeta  \right)} \right|} \right|^2} = \frac{1}{2}\mathop \sum \limits_{i = 1}^4 {\varphi ^2}\left( {{\zeta _i}} \right)$. Note that any root ${\zeta ^*}$ of $\varphi \left( \zeta  \right) = 0$ satisfies $f\left( {{\zeta ^*}} \right) = 0$. Since $f\left( \zeta  \right)$ is a positive function, every root is a minimizer of $f\left( \zeta  \right)$. However, a local minimizer of $f\left( \zeta  \right)$ is not the root of $\varphi \left( \zeta  \right) = 0$ if $f\left( \zeta  \right)$ is positive at the local minimizer. Despite this fact, the merit function is successfully used for solving a system of nonlinear equations and has been implemented in some software packages. One may directly use the MATLAB function \textit{fsolve} for this purpose, which minimizes $f\left( \zeta  \right)$ through the Gauss-Newton method that can be realized as a modified Newton's method with line search instead of solving $\varphi \left( \zeta  \right) = 0$ in a straight forward way (\cite{gratton2007approximate}). The implementation of this function for the nonlinear system Eqs. (\ref{eq17})-(\ref{eq20}) is summarized in the `Results and Discussion' section. 
\par
It may be noted that to determine the Lagrange multipliers and the entropy index by solving the system of equations, expressions for the momentum and energy coefficients are needed. For that purpose, we considered two sets of formulae available in the literature: one is based on the deterministic approach given by Chow (\cite{chow1959open}), and the other is based on Shannon entropy proposed by Chiu and Hsu (\cite{chiu2006probabilistic})
\begin{equation}\label{eq25}
    \beta  = 1 + {R_0}^2
\end{equation}
\begin{equation}\label{eq26}
    \alpha  = 1 + 3{R_0}^2 - 2{R_0}^3
\end{equation}
where $R_0$ is given by
\begin{equation}\label{eq27}
    {R_0} = \frac{{{u_{max}}}}{{\bar u}} - 1
\end{equation}
Formulae given by Eqs. (\ref{eq25}) and (\ref{eq26}) were obtained considering a logarithmic velocity profile (\cite{prandtl19257}). On the other hand, Chiu and Hsu (\cite{chiu2006probabilistic}) derived the following equations based on the velocity distribution obtained using Shannon entropy
\begin{equation}\label{eq28}
    \beta  = \frac{{\left( {{\rm{exp}}\left( {{M_c}} \right) - 1} \right)\left[ {\left( {{M_c}^2 - 2{M_c} + 2} \right){\rm{exp}}\left( {{M_c}} \right) - 2} \right]}}{{{{\left[ {\left( {{M_c} - 1} \right){\rm{exp}}\left( {{M_c}} \right) + 1} \right]}^2}}}
\end{equation}
\begin{equation}\label{eq29}
    \alpha  = \frac{{{{\left( {{\rm{exp}}\left( {{M_c}} \right) - 1} \right)}^2}\left[ {\left( {{M_c}^3 - 3{M_c}^2 + 6{M_c} - 6} \right){\rm{exp}}\left( {{M_c}} \right) + 6} \right]}}{{{{\left[ {\left( {{M_c} - 1} \right){\rm{exp}}\left( {{M_c}} \right) + 1} \right]}^3}}}
\end{equation}
where the entropy parameter $M_c$ was given implicitly as
\begin{equation}\label{eq30}
    \frac{{\bar u}}{{{u_{max}}}} = \frac{{{\rm{exp}}\left( {{M_c}} \right)}}{{{\rm{exp}}\left( {{M_c}} \right) - 1}} - \frac{1}{{{M_c}}}
\end{equation}

\subsection{Analytical Solution for Velocity Equation}
The governing differential equation for the streamwise velocity profile contains a strong nonlinear term as can be seen from Eq. (\ref{eq14}). For deriving the analytical solution of velocity equation, one may think of approximating the nonlinear term using Taylor series expansion; however, it may not produce accurate results, as the approximation will depend on the order of magnitude of the Lagrange multipliers as well as the entropy index. To that end, the Pad{\'e} approximant technique, which is considered to be the most accurate approximation of a function by a rational function of a given order, can be used (\cite{baker1996pade}). This technique often produces a better approximation of a function than its Taylor series. The $[m,n]$ order Pad{\'e} approximant of the non-linear term can be given as:
\begin{equation}\label{eq31}
    {\left( {\frac{{{q_0} - 1}}{{{q_0}}}\left[ {\frac{1}{{{q_0} - 1}} + \mathop \sum \limits_{i = 0}^2 {\lambda _i}{{\hat u}^i}} \right]} \right)^{\frac{1}{{1 - {q_0}}}}} \approx \frac{{\mathop \sum \nolimits_{i = 0}^m {C_i}{{\hat u}^i}}}{{1 + \mathop \sum \nolimits_{j = 1}^n {D_j}{{\hat u}^j}}}
\end{equation}
where $C_i$'s and $D_j$'s are the constants involving the Lagrange multipliers and the entropy index. The constants can be determined equating the like power terms on both sides. As an example, the $[2,2]$ order Pad{\'e} approximant is calculated and can be written as follows
\begin{equation}\label{eq32}
    {\left( {\frac{{{q_0} - 1}}{{{q_0}}}\left[ {\frac{1}{{{q_0} - 1}} + \mathop \sum \limits_{i = 0}^2 {\lambda _i}{{\hat u}^i}} \right]} \right)^{\frac{1}{{1 - {q_0}}}}} \approx \frac{{{C_0} + {C_1}\hat u}}{{1 + {D_1}\hat u}}
\end{equation}
where the constants are obtained as
\begin{equation}\label{eq33}
    {C_0} = {\left( {\frac{{1 - {\lambda _0} + {q_0}{\lambda _0}}}{{{q_0}}}} \right)^{\frac{1}{{1 - {q_0}}}}}
\end{equation}
\begin{equation}\label{eq34}
    {C_1} = {\left( {\frac{{1 - {\lambda _0} + {q_0}{\lambda _1}}}{{{q_0}}}} \right)^{\frac{1}{{1 - {q_0}}}}}\left[ {\frac{{2{\lambda _0}{\lambda _2} - 2{\lambda _2} - 2{\lambda _1}^2 + {q_0}\left( {{\lambda _1}^2 - 2{\lambda _0}{\lambda _2}} \right)}}{{2{\lambda _1}\left( {1 - {\lambda _0} + {\lambda _0}{q_0}} \right)}}} \right]
\end{equation}
\begin{equation}\label{eq35}
    {D_1} = \frac{{2{\lambda _0}{\lambda _2} - 2{\lambda _2} + {\lambda _1}^2{q_0} - 2{\lambda _0}{\lambda _2}{q_0}}}{{2{\lambda _1}\left( {1 - {\lambda _0} + {\lambda _0}{q_0}} \right)}}
\end{equation}
Similarly, the higher-order Pad{\'e} approximants can be obtained. Using the approximation Eq. (\ref{eq31}) and rearranging, the governing equation (\ref{eq14}) now becomes 
\begin{equation}\label{eq36}
    \left( {1 + \mathop \sum \limits_{j = 1}^n {D_j}{{\hat u}^j}} \right)\frac{{d\hat u}}{{d\hat y}} - \mathop \sum \limits_{i = 0}^m {C_i}{\hat u^i} = 0\;\;\;{\rm{subject\;to}}\;\;\hat u\left( {\hat y = 0} \right) = 0
\end{equation}
It may be noted that the Pad{\'e} approximant converts the strongly nonlinear original differential equation to a relatively weaker nonlinear form. To be specific, Eq. (\ref{eq36}) now contains integer-power nonlinearity and can be solved analytically by a non-perturbation method called the homotopy analysis method (HAM). The theoretical foundation of HAM is based on the concept of homotopy from topology to generate a convergent series solution for nonlinear ODE/PDEs (single or system). Since its inception (\cite{liao1992proposed}), the method is shown to be a unified one which logically contains most of the existing analytical methods, such as classical perturbation method, Adomian decomposition method, Lyapunov’s small artificial parameter method, the Euler transform, etc. as special cases (\cite{liao2012homotopy}). Specifically, HAM distinguishes itself from the other analytical approaches in three particular aspects: (i) its applicability is not directly confined to the presence of small physical parameters in the governing equation and/or boundary conditions; (ii) it assures the convergence of a non-linear differential equation in an efficient way through some convergence-control parameters; and (iii) it has flexibility regarding the choice of base functions and the auxiliary linear operator of the homotopy. The methodology is described briefly concerning Eq. (\ref{eq36}) in the following. 
\par
Let us write the governing Eq. (\ref{eq36}) as follows
\begin{equation}\label{eq37}
    \mathcal{N}\left[ {\hat u\left( {\hat y} \right)} \right] = 0\;\;\rm{subject\;\;to}\;\;\hat{u}\left( 0 \right) = 0
\end{equation}
where $\mathcal{N}$ is a non-linear operator or the operator of the original equation, $\hat u\left( {\hat y} \right)$ is the unknown function, and $\hat y$ is the independent variable. Now, the so-called zeroth-order deformation equation can be constructed as follows
\begin{equation}\label{eq38}
   \resizebox{1\hsize}{!}{$\mathcal{H}\left( \Phi \left( {\hat y;q} \right);q,\hbar,H \right) = (1-q) \mathcal{L}[\Phi \left( {\hat y;q} \right) - {\hat u_0}\left( {\hat y} \right)] - q \hbar H(\hat{y}) \mathcal{N}[\Phi \left( {\hat y;q} \right)] = 0\;\;\rm{subject\;to}\;\;\Phi \left( {0;q} \right)=0$}
\end{equation}
where $\mathcal{H}\left( . \right)$ is the homotopy function, $q \in \left[ {0,1} \right]$ is the embedding-parameter, $\hbar$ is a non-zero auxiliary parameter, $\mathcal{L}$ is a linear operator, and $H\left( {{\rm{\;}}\hat y} \right)$ is a non-zero auxiliary function. The core idea of HAM is that a continuous mapping is described to relate the solution $\hat u\left( {\hat y} \right)$ and the unknown function $\Phi \left( {\hat y;q} \right)$, with the aid of the embedding parameter $q$. Mathematically, as $q$ varies from $0$ to $1$, $\Phi \left( {\hat y;q} \right)$ varies from the initial approximation ${\hat u_0}\left( {\hat y} \right)$ to the final solution $\hat u\left( {\hat y} \right)$, i.e., at $q = 0$, $\Phi \left( {\hat y;q} \right) = {\rm{\;}}{\hat u_0}\left( {\hat y} \right)$ and at $q = 1$, $\Phi \left( {\hat y;q} \right) = \hat u\left( {\hat y} \right)$. It may be noted that unlike a perturbation approach (or similar other analytical methods) HAM is not directly dependent on a small parameter present in the governing equation and/or boundary conditions; here, the homotopy parameter $q$ plays the role of a small parameter. One can now define
\begin{equation}\label{eq39}
    {\hat u_m}\left( {\hat y} \right) = {\left.{\frac{1}{{m!}}\frac{{{\partial ^m}\Phi \left( {\hat y;q} \right)}}{{\partial {q^m}}}} \right|_{q = 0}}
\end{equation}
where ${\left. {\frac{{{\partial ^m}\Phi \left( {\hat y;q} \right)}}{{\partial {q^m}}}} \right|_{q = 0}}$ is called the $m$-th order deformation derivative. Using Maclaurin series expansion, one can expand $\Phi \left( {\hat y;q} \right)$ in a power series with respect to the embedding parameter $q$ as follows
\begin{equation}\label{eq40}
    \Phi \left( {\hat y;q} \right) = \Phi \left( {\hat y;0} \right) + \mathop \sum \limits_{m = 1}^\infty  {\left.{\frac{1}{{m!}}\frac{{{\partial ^m}\Phi \left( {\hat y;q} \right)}}{{\partial {q^m}}}} \right|_{q = 0}}{q^m}
\end{equation}
Assume that $\mathcal{L}$, $H\left( {\hat y} \right)$, ${\rm{\;}}{\hat u_0}\left( {\hat y} \right)$, and $\hbar$ are so properly chosen that the series Eq. (\ref{eq40}) converges at $q = 1$. Then, at $q =1$, the series becomes
\begin{equation}\label{eq41}
    \hat u\left( {\hat y} \right) = {\rm{\;}}{\hat u_0}\left( {\hat y} \right) + \mathop \sum \limits_{m = 1}^\infty  {\rm{\;}}{\hat u_m}\left( {\hat y} \right)
\end{equation}
Eq. (\ref{eq41}) provides the explicit relationship between the initial approximation ${\rm{\;}}{\hat u_0}\left( {\hat y} \right)$ and the final solution ${\rm{\;}}\hat u\left( {\hat y} \right)$. However, to obtain the explicit solution, the higher-order approximations ${\rm{\;}}{\hat u_m}\left( {\hat y} \right)$ for $m \ge 1$ need to be determined. Differentiating the zeroth-order deformation equation $m$ times with respect to $q$, and setting $q = 0$, and then dividing by $m!$, the higher-order approximations ${\rm{\;}}{\hat u_m}\left( {\hat y} \right)$ can be obtained as follows
\begin{equation}\label{eq42}
    \mathcal{L}\left[{\hat u}_{m} (\hat{y})-\chi_{m} {\hat u}_{m-1} (\hat{y})\right] =\hbar H(\hat y) R_{m}\left(\overrightarrow{\hat u}_{m-1}\right)~\text{subject to}~{\hat u}_{m}=0
\end{equation}
where
\begin{equation}\label{eq43}
    {\chi _m} = \left\{ {\begin{array}{*{20}{c}}{\;\;0\;\rm{when}\;m = 1,}\\{1\;\rm{otherwise}}\end{array}} \right.
\end{equation}
and
\begin{equation}\label{eq44}
    {R_m}\left( {{{\overrightarrow {\hat u} }_{m - 1}}} \right) = \frac{1}{{\left( {m - 1} \right)!}}{\left.{\frac{{{\partial ^{m - 1}}\mathcal{N}\left[ {\Phi \left( {\hat y;q} \right)} \right]}}{{\partial {q^{m - 1}}}}} \right|_{q = 0}}
\end{equation}
It can be seen from Eq. (\ref{eq42}) that the HAM converts the original non-linear equation into an infinite system of linear equations. The convergence of the series solution based on HAM depends on the choice of the initial approximation, linear operator, and the auxiliary function. Liao (\cite{liao2012homotopy}) proposed generalized \textit{rule of solution expression}, \textit{rule of coefficient ergodicity}, and \textit{rule of solution existence} to choose the functions appropriately. Following the theory and the rules mentioned above, the HAM-based analytical solution for Eq. (\ref{eq36}) is derived here.
\par
For solving the governing equation (\ref{eq36}), we choose the base functions $$\left\{ {{{\hat y}^m}exp\left[ {n{C_1}\hat y} \right]{\rm{|}}m,\;n = 0,1,2,3, \ldots } \right\}$$ to represent the solution $\hat u\left( {\hat y} \right)$, i.e., 
\begin{equation}\label{eq45}
    \hat u\left( {\hat y} \right) = \mathop \sum \limits_{m = 0}^\infty  \mathop \sum \limits_{n = 0}^\infty  {b_{m,n}}{\hat y^m}{\rm{exp}}\left[ {n{C_1}\hat y} \right]
\end{equation}
where $b_{m,n}$ are the coefficients. Eq. (\ref{eq45}) provides us with the so-called rule of solution expression. Accordingly, the nonlinear and linear operators are chosen as
\begin{equation}\label{eq46}
    \mathcal{N}\left[ {\Phi \left( {\hat y;q} \right)} \right] = \left[ {1 + \mathop \sum \limits_{j = 1}^n {D_j}{\Phi ^j}\left( {\hat y;q} \right)} \right]\frac{{\partial \Phi \left( {\hat y;q} \right)}}{{\partial \hat y}} - \mathop \sum \limits_{i = 0}^m {C_i}{\Phi ^i}\left( {\hat y;q} \right)
\end{equation}
\begin{equation}\label{eq47}
    \mathcal{L}\left[ {\Phi \left( {\hat y;q} \right)} \right] = \omega \frac{{\partial \Phi \left( {\hat y;q} \right)}}{{\partial \hat y}}
\end{equation}
where $\omega$ is an additional convergence-control parameter. Using Eq. (\ref{eq46}), for $\left[ {4,4} \right]$ order Pad{\'e} approximation, ${R_m}\left( {{{\overrightarrow {\hat u} }_{m - 1}}} \right)$ can be obtained from Eq. (\ref{eq44}) as follows
\begin{align}\label{eq48}
    & {R_m}\left( {{{\overrightarrow {\hat u} }_{m - 1}}} \right) = \hat u{'_{m - 1}} + {D_1}\mathop \sum \limits_{j = 0}^{m - 1} {\hat u_j}\hat u{'_{m - 1 - j}} + {D_2}\mathop \sum \limits_{j = 0}^{m - 1} {\hat u_{m - 1 - j}}\mathop \sum \limits_{k = 0}^j {\hat u_k}\hat u{'_{j - k}} \nonumber \\& + {D_3}\mathop \sum \limits_{j = 0}^{m - 1} {\hat u_{m - 1 - j}}\mathop \sum \limits_{k = 0}^j {\hat u_{j - k}}\mathop \sum \limits_{l = 0}^k {\hat u_l}\hat u{'_{k - l}} + {D_4}\mathop \sum \limits_{j = 0}^{m - 1} {\hat u_{m - 1 - j}}\mathop \sum \limits_{k = 0}^j {\hat u_{j - k}}\mathop \sum \limits_{l = 0}^n {\hat u_{n - l}}\mathop \sum \limits_{p = 0}^l {\hat u_p}\hat u{'_{l - p}} \nonumber \\& - {C_0}\left( {1 - {\chi _m}} \right) - {C_1}{\hat u_{m - 1}} - {C_2}\mathop \sum \limits_{j = 0}^{m - 1} {\hat u_j}{\hat u_{m - 1 - j}} - {C_3}\mathop \sum \limits_{j = 0}^{m - 1} {\hat u_{m - 1 - j}}\mathop \sum \limits_{k = 0}^j {\hat u_k}{\hat u_{j - k}} \nonumber \\& - {C_4}\mathop \sum \limits_{j = 0}^{m - 1} {\hat u_{m - 1 - j}}\mathop \sum \limits_{k = 0}^j {\hat u_{j - k}}\mathop \sum \limits_{l = 0}^k {\hat u_l}{\hat u_{k - l}} 
\end{align}
Applying the inverse of the linear operator Eq. (\ref{eq47}) and using the given boundary condition, the higher-order terms can be calculated from Eq. (\ref{eq42}) by the following relation
\begin{equation}\label{eq49}
    {\hat u_m}\left( {\hat y} \right) = {\chi _m}{\hat u_{m - 1}}\left( {\hat y} \right) + \frac{\hbar }{\omega }\mathop \int \limits_0^{\hat y} H\left( {\hat y} \right){R_m}\left( {{{\overrightarrow {\hat u} }_{m - 1}}} \right)d\hat y
\end{equation}
where $R_m$'s are given by Eq. (\ref{eq48}). It may be noted that the auxiliary function is selected as $H(\hat{y})=$1 to avoid computational difficulty. This can always be done if one selects an optimal linear operator (\cite{vajravelu2013nonlinear}). Finally, the $M$-th order HAM-based approximate analytical solution can be obtained explicitly as
\begin{equation}\label{eq50}
    \hat u\left( {\hat y} \right) \approx {\hat u_{SUM}}\left( {\hat y} \right) = \mathop \sum \limits_{j = 0}^M {\hat u_j}\left( {\hat y} \right)
\end{equation}
It can be noticed from Eq. (\ref{eq40}) that HAM itself is a kind of generalized Taylor series at $q=1$. Hence, if the initial approximation is chosen accurately enough, then one can obtain the rapid convergence of the series (\cite{nave2013comparison}). To that end, preserving the \textit{rule of solution expression} as well, the following initial approximation was selected
\begin{equation}\label{eq51}
    {\hat u_0}\left( {\hat y} \right) = \frac{{{C_0}}}{{{C_1}}}\left[ {\exp\left( {{C_1}\hat y} \right) - 1} \right]
\end{equation}
Eq. (\ref{eq51}) is indeed the solution of the part of the original equation. For the sake of completeness, the theoretical convergence analysis for the series solution Eq. (\ref{eq50}) is shown in \textit{Appendix A}. 

\section{Results and Discussion}
The velocity model developed above requires the determination of entropy index, momentum and energy coefficients, and Lagrange multipliers which is done using experimental and field data. Then, the HAM-based approximate analytical solution is validated with the numerical solution. Finally, the derived velocity model results are compared with experimental and field data and also with the existing equation based on Tsallis entropy. Each of these steps is discussed in what follows. 

\subsection{Laboratory and Field Data}
For validating the derived velocity profile, relevant sets of experimental and field data from the literature were utilized. The present study considers the vertical distribution of streamwise velocity, where the velocity increases monotonically from the channel bed, and the maximum velocity appears at the water surface. To that end, the most cited laboratory data of Vanoni (\cite{vanoni1946transportation}) and Einstein and Chien (\cite{einstein1955effects}) were chosen, while the field data were collected from Davoren (\cite{davoren1985local}). The velocity data of Vanoni (\cite{vanoni1946transportation}) were collected from the experiments carried out in two series: clear water and sediment-mixed fluid. The aspect ratio of the experimental channel varied from 5 to 11.9 and maximum velocity appeared at the free surface. On the other hand, Einstein and Chien (\cite{einstein1955effects}) conducted near-bed (up to 50\% of the flow depth starting from channel bed) experiments in a painted steel flume for both clear water and sediment-laden flow. The sediment-mixed fluid experiment was having sediment particles with three different sizes. Further details are available in \cite{einstein1955effects}.  Field data of Davoren (\cite{davoren1985local}) was for a river with a live bed. The flow velocities were measured downstream from a hydropower plant, which resulted in steady uniform flow over an appreciable period of time. The relevant flow parameters for some selected data sets are reported in Table \ref{t1}. 
\begin{table}[]
\centering
\caption{Characteristics of the selected data sets.}\label{t1}
\scalebox{0.7}{\begin{tabular}{|c|c|c|c|c|c|c|c|c|}
\hline
\multicolumn{2}{|c|}{\multirow{2}{*}{Data Set}} & \multirow{2}{*}{\begin{tabular}[c]{@{}c@{}}Mean \\ velocity \\ $\bar{u}$ (m/s)\end{tabular}} & \multirow{2}{*}{\begin{tabular}[c]{@{}c@{}}Maximum \\ velocity \\ $u_{max}$ (m/s)\end{tabular}} & \multirow{2}{*}{\begin{tabular}[c]{@{}c@{}}Entropy \\ parameter\\  $M_{c}$\end{tabular}} & \multicolumn{2}{c|}{\begin{tabular}[c]{@{}c@{}}Momentum\\  coefficient\\  $\beta$\end{tabular}} & \multicolumn{2}{c|}{\begin{tabular}[c]{@{}c@{}}Energy\\  coefficient \\ $\alpha$\end{tabular}} \\ \cline{6-9} 
\multicolumn{2}{|c|}{} &  &  &  & Eq. (\ref{eq25}) & Eq. (\ref{eq28}) & Eq. (\ref{eq26}) & Eq. (\ref{eq29}) \\ \hline
\multirow{2}{*}{Vanoni (\cite{vanoni1946transportation})} & Run 15 & 1.153 & 1.360 & 6.5060 & 1.0322 & 1.0308 & 1.0851 & 1.0829 \\ \cline{2-9} 
 & Run 20 & 0.941 & 1.176 & 4.8059 & 1.0624 & 1.0546 & 1.1560 & 1.1452 \\ \hline
\multirow{2}{*}{\begin{tabular}[c]{@{}c@{}}Einstein and \\ Chien (\cite{einstein1955effects})\end{tabular}} & Run C3 & 1.960 & 2.288 & 6.9280 & 1.0280 & 1.0271 & 1.0746 & 1.0732 \\ \cline{2-9} 
 & Run S5 & 2.226 & 2.802 & 4.6454 & 1.0670 & 1.0579 & 1.1662 & 1.1538 \\ \hline
\multirow{2}{*}{Davoren (\cite{davoren1985local})} & Run 1 & 2.285 & 2.807 & 5.2260 & 1.0522 & 1.0471 & 1.1327 & 1.1254 \\ \cline{2-9} 
 & Run 10 & 2.258 & 2.790 & 5.0770 & 1.0560 & 1.0500 & 1.1400 & 1.1320 \\ \hline
\end{tabular}}
\end{table}
\subsection{Validation of the HAM-Based Approximate Series Solution}
The HAM-based series solution Eq. (\ref{eq50}) was validated by comparing it with the numerical solution for a relevant set of data. It can be observed that the series solution based on HAM depends on two convergence-control parameters, namely ${\omega}$ and $\hbar$, which need to be determined for assessing the solution. In the framework of HAM, the convergence control parameters play one of the most vital roles. A suitable choice for the parameters leads the series solution towards the exact solution over the entire domain. Also, unlike the other perturbation and non-perturbation methods, these convergence-control parameters greatly enhance the radius and rate of convergence of the series solution (\cite{liao2012homotopy}). For the determination of parameters, the squared residual error ($\Delta_{m}$) method was used, in which $\Delta_{m}$ at ${m}$-th order approximation reads as
\begin{equation}\label{eq52}
   \Delta_{m} = \int_{\hat{y} \in \Omega} (\mathcal{N}[\hat{u}(\hat{y})])^{2} d\hat{y}
\end{equation}
Here $\Omega$ is the domain which is $[0,1]$ for the present problem. In this method, for a particular order of approximation $m$, the corresponding $\Delta_{m}$ is minimized (i.e., $\frac{\partial \Delta_{m}}{\partial \omega} = 0$ and $\frac{\partial \Delta_{m}}{\partial \hbar} = 0$) to get two equations with two unknowns $\hbar$ and $\omega$ which determine the optimum values of convergence parameters. However, due to the analytical integration, Eq. (\ref{eq52}) may sometime create a computational difficulty. To avoid this difficulty, the discrete squared residual error, ${E_m}$, is often calculated as follows
\begin{equation}\label{eq53}
   E_{m} = \frac{1}{L+1} \sum_{j=0}^{L} \Big[\mathcal{N}\Big(\sum_{k=0}^{m} {\hat{u}_{k}} (j\Delta \hat{y})\Big)\Big]^{2}  
\end{equation}
where $L+1$ are the equally distributed discrete points. The homotopy series solution leads to the exact solution when the squared residual error tends to zero. Therefore, for the convergence of solution, it is sufficient to check the residual error Eq. (\ref{eq52}) or (\ref{eq53}) only.
\par
\begin{figure}[]
       \centering
       \begin{subfigure}[b]{0.21\textwidth}           
           \includegraphics[scale=0.31]{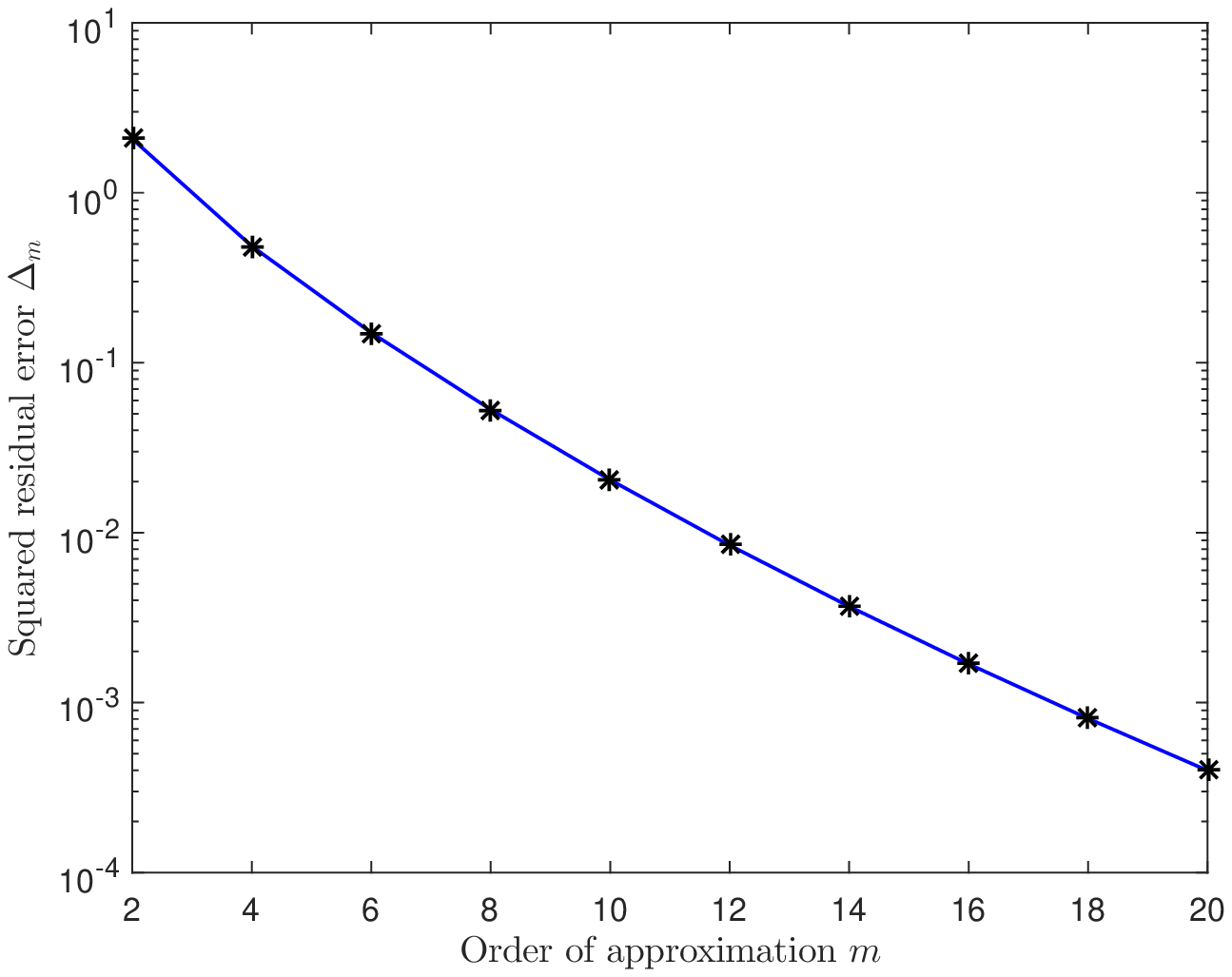}           
                \caption{}
                \label{fig1a}
       \end{subfigure}%
       \hspace{1.5cm}
              \begin{subfigure}[b]{0.21\textwidth}           
                \includegraphics[scale=0.31]{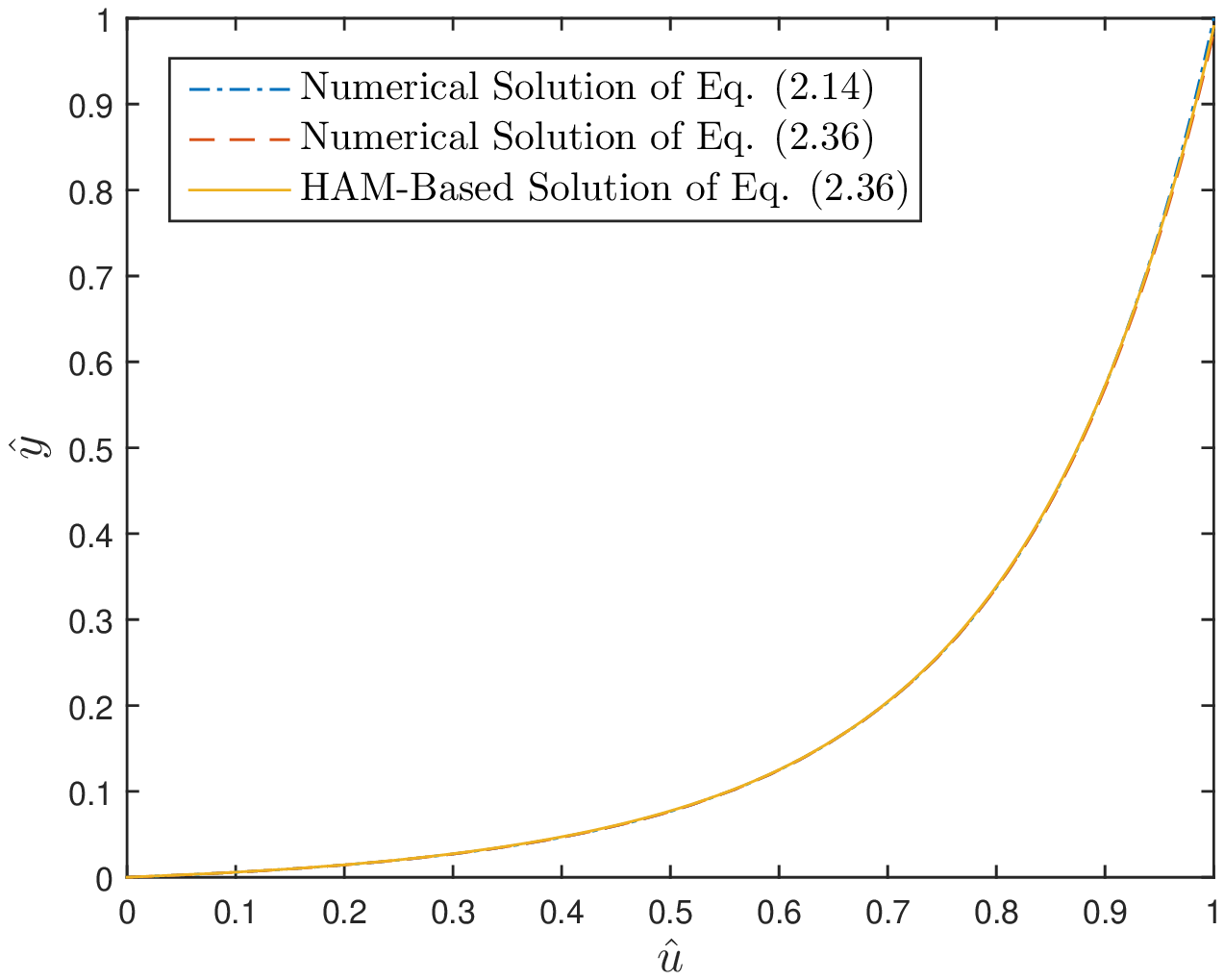}               
                \caption{}
                \label{fig1b}
       \end{subfigure}%
 \caption{Validation of the approximated series solution: (a) Squared residual error with different order of approximations of HAM-based series solution and (b) Comparison of $20^{th}$ order HAM-based analytical solution with numerical solutions of Eqs. (\ref{eq14}) and (\ref{eq36}).}\label{fig1}
\end{figure}

Now, a test case is performed to check the validity of the HAM-based approximate analytical solution. For that purpose, $[4,4]$ order Pad{\'e} approximation is considered where the values of the parameters are $\lambda_0=-2.2077$, $\lambda_1=4.5346$, $\lambda_2=0.0852$, and $q_{0}=0.8873$, and the Pad{\'e} approximation coefficients are $C_0=20.7488$, $C_1=-56.1968$, $C_2=60.0071$, $C_3=-30.3244$, $C_4=6.2342$, $D_1=0.9227$, $D_2=0.4612$, $D_3=0.1431$, and $D_4=0.0232$.  With these parameters, the square residual errors given by Eq. (\ref{eq52}) are plotted in Fig. \ref{fig1a} with different orders of approximations of HAM-based solution. It is seen from the figure that the residual error $\Delta_m$ decreases systematically with the increasing order of approximation $m$, and hence the choice of operators and initial approximation is suitable to guarantee the convergence pattern of the method. 
\par
Next, the HAM-based solution is verified with the numerical solutions of Eqs. (\ref{eq14}) and (\ref{eq36}). The original governing equation Eq. (\ref{eq14}) was converted to a relatively weaker non-linear equation Eq. (\ref{eq36}) using Pad{\'e} approximation. Therefore, solutions of both equations are compared with the HAM-based series solution in order to get a comparative idea as well as the validity of the approximation made. The function \textit{ode45} of MATLAB was used for the numerical solution. The convergence-control parameters were calculated using the squared residual error method described in the aforementioned discussion. The $20^{th}$ order HAM-based approximation along with numerical solutions are plotted in Fig. \ref{fig1b}. Apart from the graphical presentation, a quantitative assessment is shown in Table 2 for some discrete points within the domain. It can be observed from the figure and the table that the HAM-based approximate solution Eq. (\ref{eq50}) is close to the numerical solution of Eq. (\ref{eq36}) obtained via $[4,4]$ Pad{\'e} approximation. On the other hand, the solution of the original governing equation Eq. (\ref{eq14}) slightly differs from the solution of the approximated equation at the right end points of the domain. This deviation is attributed to the order of the Pad{\'e} approximation; indeed, further accuracy can be achieved if one sought for higher-order approximation. Overall, the $[4,4]$ Pad{\'e} approximation is seen to be an adequate approximation for the nonlinear term. Also, for engineering applications, the time taken by the computer for some specific order of approximation is given in Table 3. It can be seen from the table that the methodology can be performed efficiently without time complexity if a simple (proper) set of base function is selected.

\begin{table}[!htb]
\caption{Numerical comparison between HAM-based approximation and numerical solution.}\label{t2}
\centering
\begin{tabular}{|c|c|c|c|c|}
\hline
\multirow{2}{*}{$\hat{y}$} & \multirow{2}{*}{\begin{tabular}[c]{@{}c@{}}Numerical\\ solution\\ Eq. (\ref{eq36})\end{tabular}} & \multicolumn{2}{c|}{\begin{tabular}[c]{@{}c@{}}HAM-based\\ approximation\\ Eq. (\ref{eq50})\end{tabular}} & \multirow{2}{*}{\begin{tabular}[c]{@{}c@{}}Numerical\\ solution\\ Eq. (\ref{eq14})\end{tabular}} \\ \cline{3-4}
 &  & $10^{th}$ order & $20^{th}$ order &  \\ \hline
0 & 0.0000 & 0.0000 & 0.0000 & 0.0000 \\ \hline
0.1 & 0.5547 & 0.5426 & 0.5534 & 0.5547 \\ \hline
0.2 & 0.6964 & 0.6927 & 0.6956 & 0.6964 \\ \hline
0.3 & 0.7772 & 0.7760 & 0.7763 & 0.7771 \\ \hline
0.4 & 0.8332 & 0.8322 & 0.8324 & 0.8329 \\ \hline
0.5 & 0.8756 & 0.8747 & 0.8748 & 0.8751 \\ \hline
0.6 & 0.9097 & 0.9087 & 0.9088 & 0.9089 \\ \hline
0.7 & 0.9382 & 0.9370 & 0.9373 & 0.9369 \\ \hline
0.8 & 0.9626 & 0.9612 & 0.9617 & 0.9608 \\ \hline
0.9 & 0.9840 & 0.9819 & 0.9832 & 0.9816 \\ \hline
1.0 & 1.0031 & 0.9948 & 1.0017 & 1.0000 \\ \hline
\end{tabular}
\end{table}

\begin{table}[!htb]
\caption{Squared residual error and computational time for different order of approximations.}\label{t3}
\centering
\begin{tabular}{|c|c|c|}
\hline
\begin{tabular}[c]{@{}c@{}}Order of\\ approximation \\ $m$\end{tabular} & \begin{tabular}[c]{@{}c@{}}Squared residual\\ error $\Delta_{m}$\end{tabular} & \begin{tabular}[c]{@{}c@{}}Time\\ (sec)\end{tabular} \\ \hline
2 & $2.07 \times 10^{+0}$ & 0.306 \\ \hline
4 & $4.85 \times 10^{-1}$ & 0.781 \\ \hline
6 & $1.50 \times 10^{-1}$ & 1.667 \\ \hline
8 & $5.30\times 10^{-2}$ & 3.467 \\ \hline
10 & $2.04 \times 10^{-2}$ & 7.842 \\ \hline
12 & $8.43 \times 10^{-3}$ & 17.970 \\ \hline
14 & $3.68 \times 10^{-3}$ & 41.822 \\ \hline
16 & $1.69 \times 10^{-3}$ & 93.270 \\ \hline
18 & $8.06 \times 10^{-4}$ & 203.834 \\ \hline
20 & $3.98 \times 10^{-4}$ & 441.858 \\ \hline
\end{tabular}
\end{table}

\subsection{Comparison with Experimental and Field Data}
The proposed velocity model was validated by comparing it with selected sets of data, namely, Run 20, S5, and 1 of Vanoni (\cite{vanoni1946transportation}), Einstein and Chien (\cite{einstein1955effects}), and Davoren (\cite{davoren1985local}), respectively. The Lagrange multipliers and the entropy index needed to assess the model were calculated by solving the system of nonlinear equations Eqs. (\ref{eq17})-(\ref{eq20}) using the Gauss-Newton method after approximating the constraints integrals using a 25-point Gauss-Legendre quadrature rule, and are reported in Table \ref{t4}. The `fsolve' command in MATLAB R2014b implements the Gauss-Newton method for the given system of nonlinear equations. Here we used the command as its simplest structure, where we put the system and the starting point. The inbuilt line search scheme in this MATLAB command relaxes the burden of choosing a starting point close to the solution. Then, the HAM-based analytical solution of Eq. (\ref{eq36}) was assessed for comparison, and both the formulae for momentum ($\beta$) and energy ($\alpha$) coefficients given by Eqs. (\ref{eq25})-(\ref{eq26}) and (\ref{eq28})-(\ref{eq29}) were considered. The convergence-control parameters were determined using the squared residual error given by Eq. (\ref{eq52}). The Pad{\'e} approximant coefficients were determined using the function `PadeApproximant' of MATHEMATICA and are reported in Table \ref{t5}.
\par
\begin{figure}[htbp]
  \centering
  \includegraphics[scale=0.4]{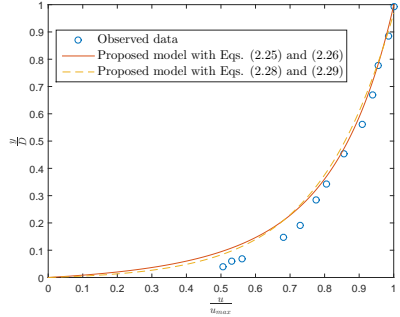}
  \caption{Comparison of the HAM-based velocity profile Eq. (\ref{eq50}) with Run 20 of Vanoni (\cite{vanoni1946transportation}) data: $20^{th}$ order approximations for the model with Eqs. (\ref{eq25})-(\ref{eq26}), and (\ref{eq28})-(\ref{eq29}).}\label{fig2}
\end{figure}
\begin{figure}[htbp] 
  \centering
  \includegraphics[scale=0.4]{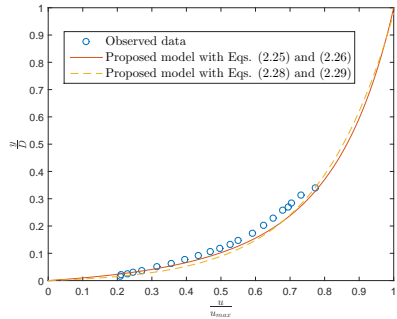}
  \caption{Comparison of the HAM-based velocity profile Eq. (\ref{eq50}) with Run S5 of Einstein and Chien (\cite{einstein1955effects}) data: $20^{th}$ and $25^{th}$ order approximations for the model with Eqs. (\ref{eq25})-(\ref{eq26}), and (\ref{eq28})-(\ref{eq29}), respectively.}\label{fig3}
\end{figure}
\begin{figure}[htbp] 
  \centering
  \includegraphics[scale=0.4]{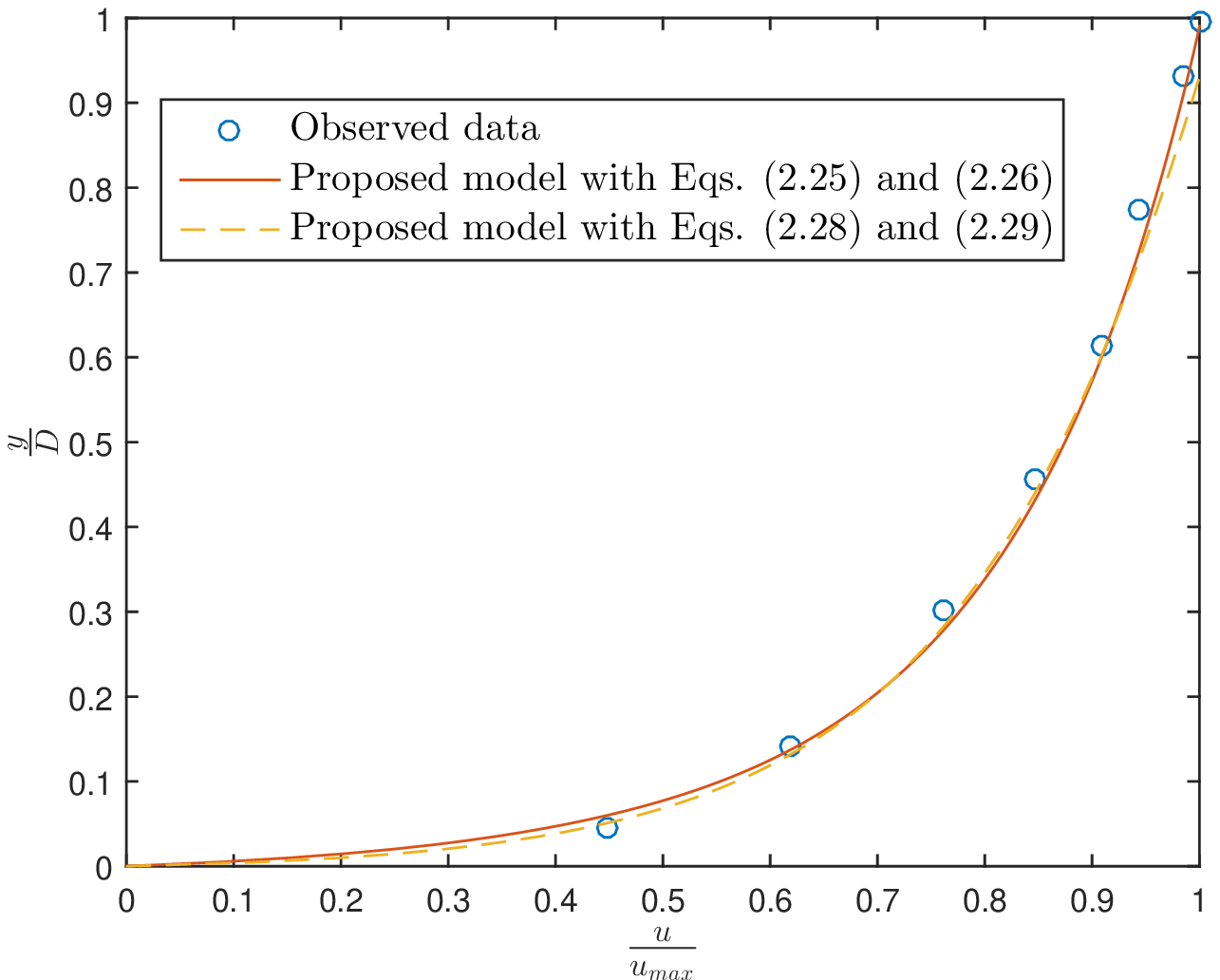}
  \caption{Comparison of the HAM-based velocity profile Eq. (\ref{eq50}) with Run 1 of Davoren (\cite{davoren1985local}) data: $20^{th}$ and $30^{th}$ order approximations for the model with Eqs. (\ref{eq25})-(\ref{eq26}), and (\ref{eq28})-(\ref{eq29}), respectively.}\label{fig4}
\end{figure}

Fig. \ref{fig2} compares the proposed model with Run 20 of Vanoni (\cite{vanoni1946transportation}) data. $20^{th}$ order HAM-based approximate solution was considered for the model with both cases of $\beta$ and $\alpha$. It can be observed from the figure that the model for the set of both $\beta$ and $\alpha$ agreed well with the experimental data, specifically after 30\% of the flow depth. The model corresponding to $\beta$ and $\alpha$ from Eqs. (\ref{eq25})-(\ref{eq26}) performed better in the upper half of the channel while the other profile dominated near the channel bed. In Fig. \ref{fig3}, a data set, namely Run S5 from the experiment of the sediment-laden flow of Einstein and Chien (\cite{einstein1955effects}) was compared with $20^{th}$ and $25^{th}$ order HAM-based approximations for the model with $\beta$ and $\alpha$ from Eqs. (\ref{eq25})-(\ref{eq26}) and (\ref{eq28})-(\ref{eq29}), respectively. It is seen from the figure that the model for both cases estimated the experimental values well, though the near-bed prediction accuracy was superior for the model with $\beta$ and $\alpha$ from Eqs. (\ref{eq25}) and (\ref{eq26}). Fig. \ref{fig4} shows the comparison of the model with field data (Run 1) from Davoren (\cite{davoren1985local}). Here $20^{th}$ and $30^{th}$ order HAM-based approximations were considered for the model with $\beta$ and $\alpha$ from Eqs. (\ref{eq25})-(\ref{eq26}) and (\ref{eq28})-(\ref{eq29}), respectively. It can be observed from the figure that the model for both cases performed well in measuring velocity throughout the water column. Overall, it can be concluded that the proposed model predicted the vertically distributed velocities in open channels well. It may be noted that the information-theoretic concept, along with the maximum entropy principle in hydraulics, does not incorporate explicit fluid mechanics processes; the physics can only be included through the constraints and the available data. 

\begin{table}[]
\centering
\caption{Values of the Lagrange multipliers and the entropy index for the velocity models for some selected sets of data.}\label{t4}
\scalebox{0.63}{\begin{tabular}{|c|c|c|c|c|c|c|c|c|c|c|c|c|}
\hline
\multicolumn{2}{|c|}{\multirow{2}{*}{\begin{tabular}[c]{@{}c@{}}Data\\ Set\end{tabular}}} & \multicolumn{4}{c|}{\begin{tabular}[c]{@{}c@{}}$\beta$ and $\alpha$ from\\ Eqs. (\ref{eq25}) and (\ref{eq26})\end{tabular}} & \multicolumn{4}{c|}{\begin{tabular}[c]{@{}c@{}}$\beta$ and $\alpha$ from\\ Eqs. (\ref{eq28}) and (\ref{eq29})\end{tabular}} & \multicolumn{3}{c|}{\begin{tabular}[c]{@{}c@{}}Singh and Luo\\ (\cite{singh2011entropy}) model\end{tabular}} \\ \cline{3-13} 
\multicolumn{2}{|c|}{} & $\lambda_{0}$ & $\lambda_{1}$ & $\lambda_{2}$ & $q_{0}$ & $\lambda_{0}$ & $\lambda_{1}$ & $\lambda_{2}$ & $q_{0}$ & $\lambda_{0}$ & $\lambda_{1}$ & $q_{0}$ \\ \hline
\multirow{2}{*}{\begin{tabular}[c]{@{}c@{}}Vanoni\\ (\cite{vanoni1946transportation})\end{tabular}} & Run 15 & -3.4825 & 6.2386 & 0.0240 & 0.9652 & -3.6251 & 6.4837 & 0.0175 & 1.0004 & -2.6237 & 4.8642 & 0.750 \\ \cline{2-13} 
 & Run 20 & -1.7380 & 3.8497 & 0.1285 & 0.8320 & -2.2176 & 4.7441 & 0.0703 & 1.0068 & -1.5575 & 3.6334 & 0.750 \\ \hline
\multirow{2}{*}{\begin{tabular}[c]{@{}c@{}}Einstein\\ and Chien (\cite{einstein1955effects})\end{tabular}} & Run C3 & -3.8746 & 6.7247 & 0.0159 & 0.9754 & -3.9820 & 6.9044 & 0.0123 & 0.9992 & -2.8854 & 5.1583 & 0.750 \\ \cline{2-13} 
 & Run S5 & -1.5459 & 3.5606 & 0.1502 & 0.8039 & -2.0876 & 4.5737 & 0.0801 & 1.0076 & -1.4571 & 3.5139 & 0.750 \\ \hline
\multirow{2}{*}{\begin{tabular}[c]{@{}c@{}}Davoren\\ (\cite{davoren1985local})\end{tabular}} & Run 1 & -2.208 & 4.535 & 0.085 & 0.8870 & -2.5540 & 5.1710 & 0.0510 & 1.0030 & -1.8209 & 3.9435 & 0.750 \\ \cline{2-13} 
 & Run 10 & -2.0149 & 4.2466 & 0.1019 & 0.8599 & -2.4085 & 4.9717 & 0.0596 & 0.9948 & -1.7277 & 3.8344 & 0.750 \\ \hline
\end{tabular}}
\end{table}

\begin{table}[]
\centering
\caption{Pad{\'e} approximant coefficients for some selected sets of data.}\label{t5}
\scalebox{0.68}{\begin{tabular}{|c|c|c|c|c|c|}
\hline
\multicolumn{2}{|c|}{\multirow{3}{*}{Data Set}} & \multicolumn{4}{c|}{Pad{\'e} approximant coefficients} \\ \cline{3-6} 
\multicolumn{2}{|c|}{} & \multicolumn{2}{c|}{$\beta$ and $\alpha$ from Eqs. (\ref{eq25}) and (\ref{eq26})} & \multicolumn{2}{c|}{$\beta$ and $\alpha$ from Eqs. (\ref{eq28}) and (\ref{eq29})} \\ \cline{3-6} 
\multicolumn{2}{|c|}{} & $(C_{0},C_{1},C_{2},C_{3},C_{4})$ & $(D_{1},D_{2},D_{3},D_{4})$ & $(C_{0},C_{1},C_{2},C_{3},C_{4})$ & $(D_{1}, D_{2}, D_{3}, D_{4})$ \\ \hline
Vanoni (\cite{vanoni1946transportation}) & Run 20 & \begin{tabular}[c]{@{}c@{}}(13.731, -35.980, 36.263,\\ -16.781, 3.041)\end{tabular} & \begin{tabular}[c]{@{}c@{}}(0.359, 0.118, 0.026,\\ 0.003)\end{tabular} & \begin{tabular}[c]{@{}c@{}}(25.307, -60.792,\\ 62.633, -33.485,\\ 8.062)\end{tabular} & \begin{tabular}[c]{@{}c@{}}(2.415, 2.497,\\ 1.339, 0.323)\end{tabular} \\ \hline
\begin{tabular}[c]{@{}c@{}}Einstein and Chien \\ (\cite{einstein1955effects})\end{tabular} & Run S5 & \begin{tabular}[c]{@{}c@{}}(11.744, -30.515, 30.195,\\ -13.546, 2.341)\end{tabular} & \begin{tabular}[c]{@{}c@{}}(0.134, 0.052, 0.009,\\ 0.001)\end{tabular} & \begin{tabular}[c]{@{}c@{}}(22.210, -51.597,\\ 51.385, -26.541,\\ 6.170)\end{tabular} & \begin{tabular}[c]{@{}c@{}}(2.324, 2.316,\\ 1.196, 0.278)\end{tabular} \\ \hline
Davoren (\cite{davoren1985local}) & Run 1 & \begin{tabular}[c]{@{}c@{}}(20.749, -56.197, 60.007,\\ -30.324, 6.234)\end{tabular} & \begin{tabular}[c]{@{}c@{}}(0.923, 0.461,\\ 0.143, 0.023)\end{tabular} & \begin{tabular}[c]{@{}c@{}}(35.245, -92.119,\\ 103.112, -59.802,\\ 15.596)\end{tabular} & \begin{tabular}[c]{@{}c@{}}(2.597, 2.894,\\ 1.673, 0.436)\end{tabular} \\ \hline
\end{tabular}}
\end{table}

\subsection{Comparison with the Existing Equation using Tsallis Entropy}
Here, we compare the proposed model with the existing model on Tsallis entropy to check the effect of additional constraints as well as the entropy index on the vertically distributed velocity. Singh and Luo (\cite{singh2011entropy}) considered only the total probability theorem and the mass conservation constraint in their derivation and also concluded a fixed value 3/4 for the entropy index based on a data-fitting procedure. For the sake of completeness, the model proposed by \cite{singh2011entropy} is given here as follows:
\begin{equation}\label{eq54}
    \hat u =  - \frac{{{\lambda _0} + \frac{1}{{{q_0} - 1}}}}{{{\lambda _1}}} + \frac{1}{{{\lambda _1}}}{\left[ {{{\left( {{\lambda _0} + \frac{1}{{{q_0} - 1}}} \right)}^{\frac{{{q_0}}}{{{q_0} - 1}}}} + {\lambda _1}{{\left( {\frac{{{q_0}}}{{{q_0} - 1}}} \right)}^{\frac{{{q_0}}}{{{q_0} - 1}}}}\hat y} \right]^{\frac{{{q_0} - 1}}{{{q_0}}}}}
\end{equation}
where the index $q_0$ was chosen as 3/4, the Lagrange multipliers were obtained by solving the constraint Eqs. (\ref{eq2}) and (\ref{eq3}). The proposed model was assessed for the set of both formulae of $\beta$ and $\alpha$ given by Eqs. (\ref{eq25}), (\ref{eq26}), (\ref{eq28}), and (\ref{eq29}). To check the prediction accuracies of each of the models, the relative error (RE) and root-mean-squared error (RMSE) were calculated as follows
\begin{equation}\label{eq55}
RE = \frac{1}{{{M_0}}}\mathop \sum \limits_{i = 1}^{{M_0}} \left| {\frac{{{{\hat u}_{obs}}\left( i \right) - {{\hat u}_{com}}\left( i \right)}}{{{{\hat u}_{obs}}\left( i \right)}}} \right|
\end{equation}
\begin{equation}\label{eq56}
RMSE = \sqrt {\frac{1}{{{M_0}}}\mathop \sum \limits_{i = 1}^{{M_0}} {{\left( {\frac{{{{\hat u}_{obs}}\left( i \right) - {{\hat u}_{com}}\left( i \right)}}{{{{\hat u}_{obs}}\left( i \right)}}} \right)}^2}}
\end{equation}
where ${\hat u_{obs}}\left( i \right)$ and ${\hat u_{com}}\left( i \right)$ are observed and computed values of the normalized velocity at the $i$-th data point, and ${M_0}$ is the total number of data points. 
\par
\begin{figure}[]
       \centering
       \begin{subfigure}[b]{0.21\textwidth}           
           \includegraphics[scale=0.31]{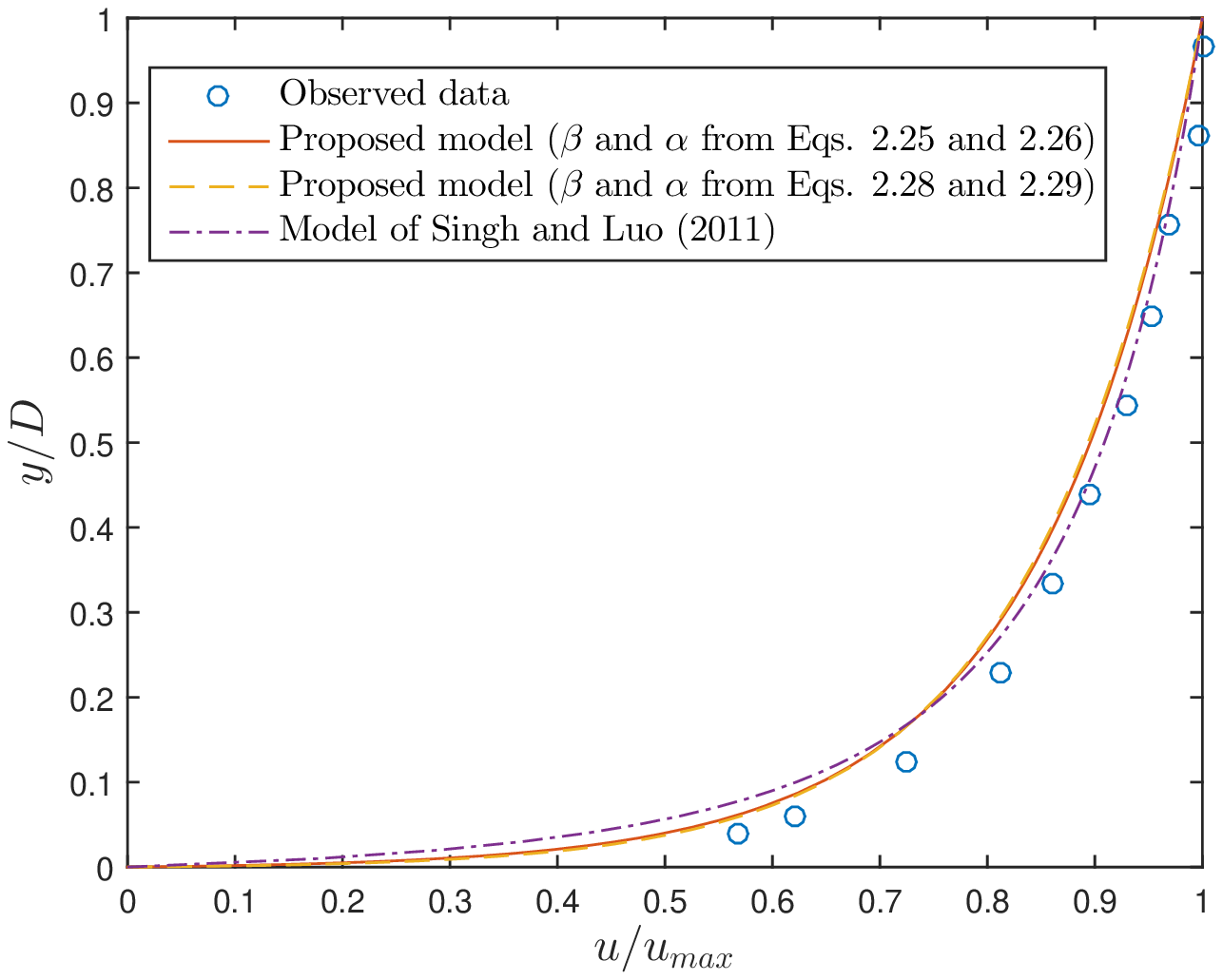}           
                \caption{}
                \label{fig5a}
       \end{subfigure}%
       \hspace{1.5cm}
              \begin{subfigure}[b]{0.21\textwidth}           
                \includegraphics[scale=0.31]{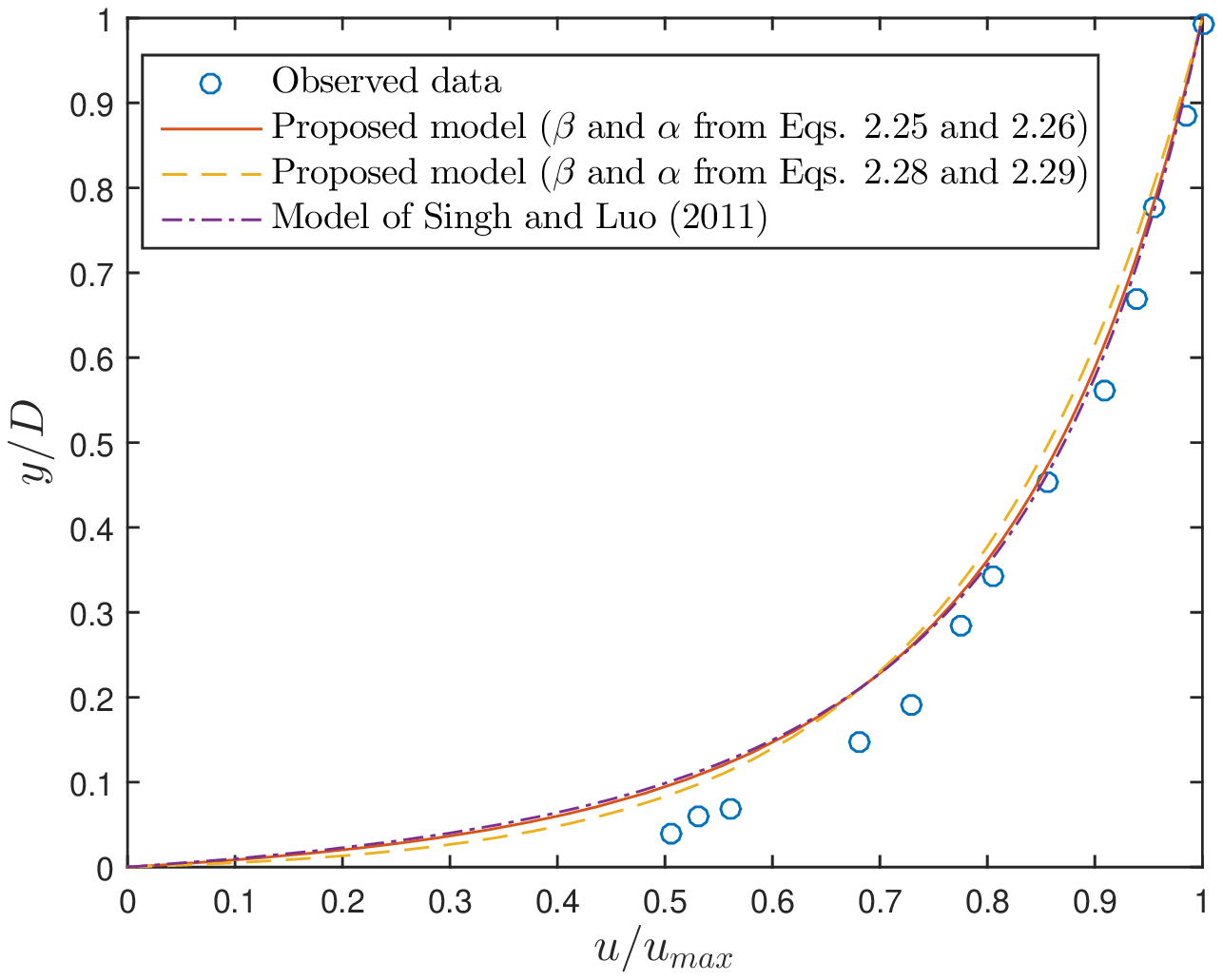}
                \caption{}
                \label{fig5b}
       \end{subfigure}%
 \caption{Comparison of the proposed model with the model of Singh and Luo (\cite{singh2011entropy}) for (a) Run 15, and (b) Run 20 of Vanoni (\cite{vanoni1946transportation}) data.}\label{fig5}
\end{figure}
\begin{figure}[]
       \centering
       \begin{subfigure}[b]{0.21\textwidth}           
           \includegraphics[scale=0.31]{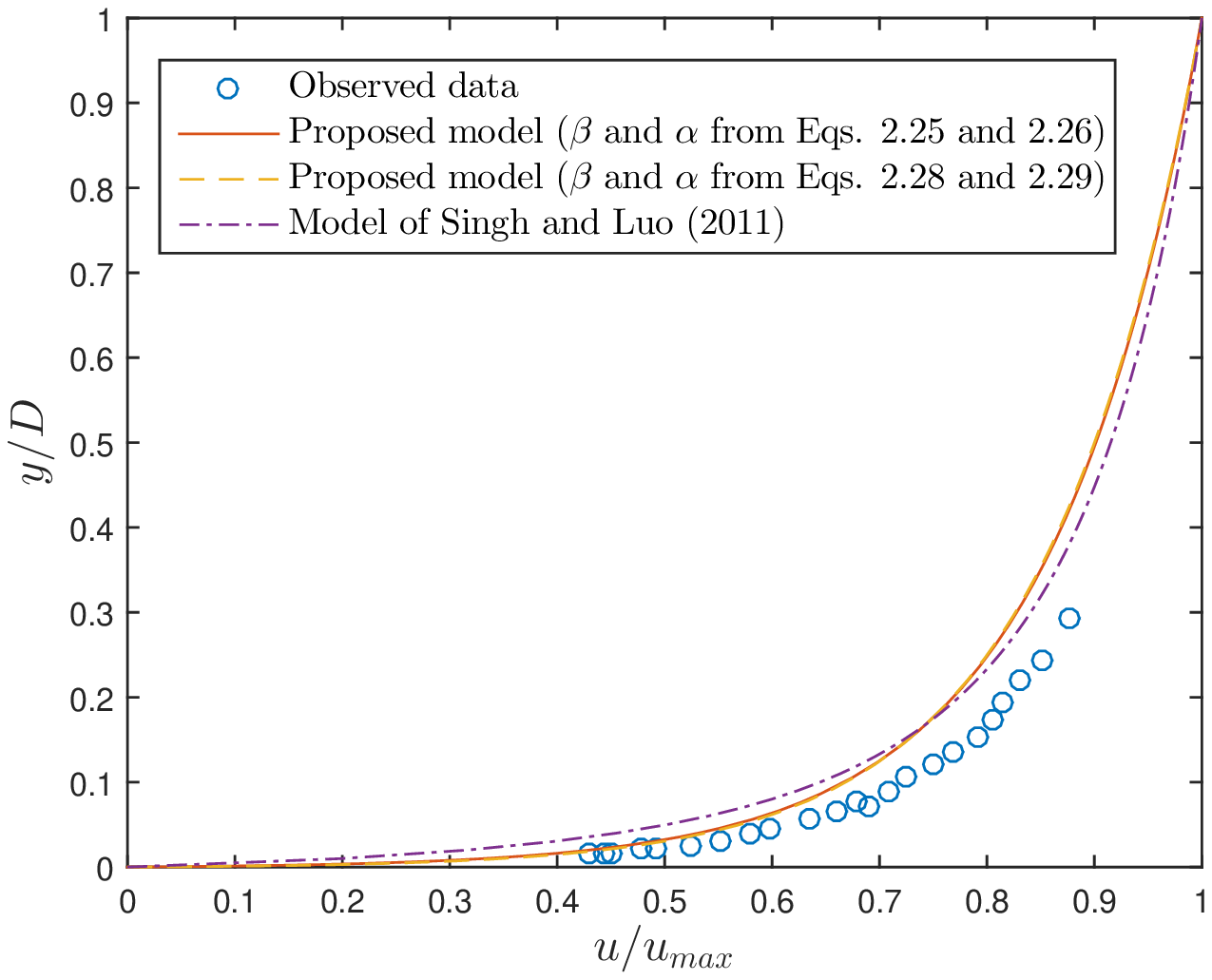}           
                \caption{}
                \label{fig6a}
       \end{subfigure}%
       \hspace{1.5cm}
              \begin{subfigure}[b]{0.21\textwidth}           
                \includegraphics[scale=0.31]{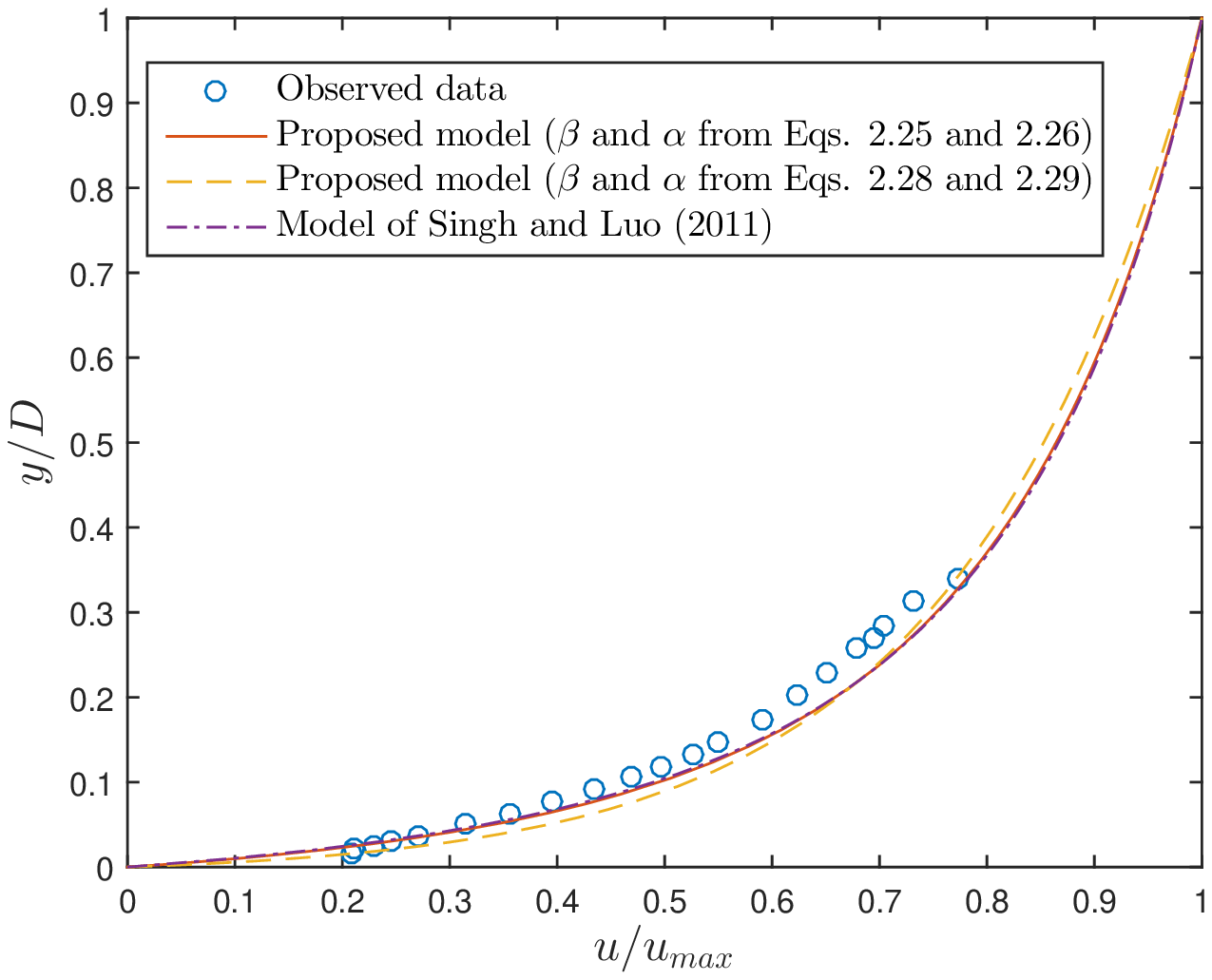}
                \caption{}
                \label{fig6b}
       \end{subfigure}%
 \caption{Comparison of the proposed model with the model of Singh and Luo (\cite{singh2011entropy}) for (a) Run C3, and (b) Run S5 of Einstein and Chien (\cite{einstein1955effects}) data.}\label{fig6}
\end{figure}
\begin{figure}[]
       \centering
       \begin{subfigure}[b]{0.21\textwidth}           
           \includegraphics[scale=0.31]{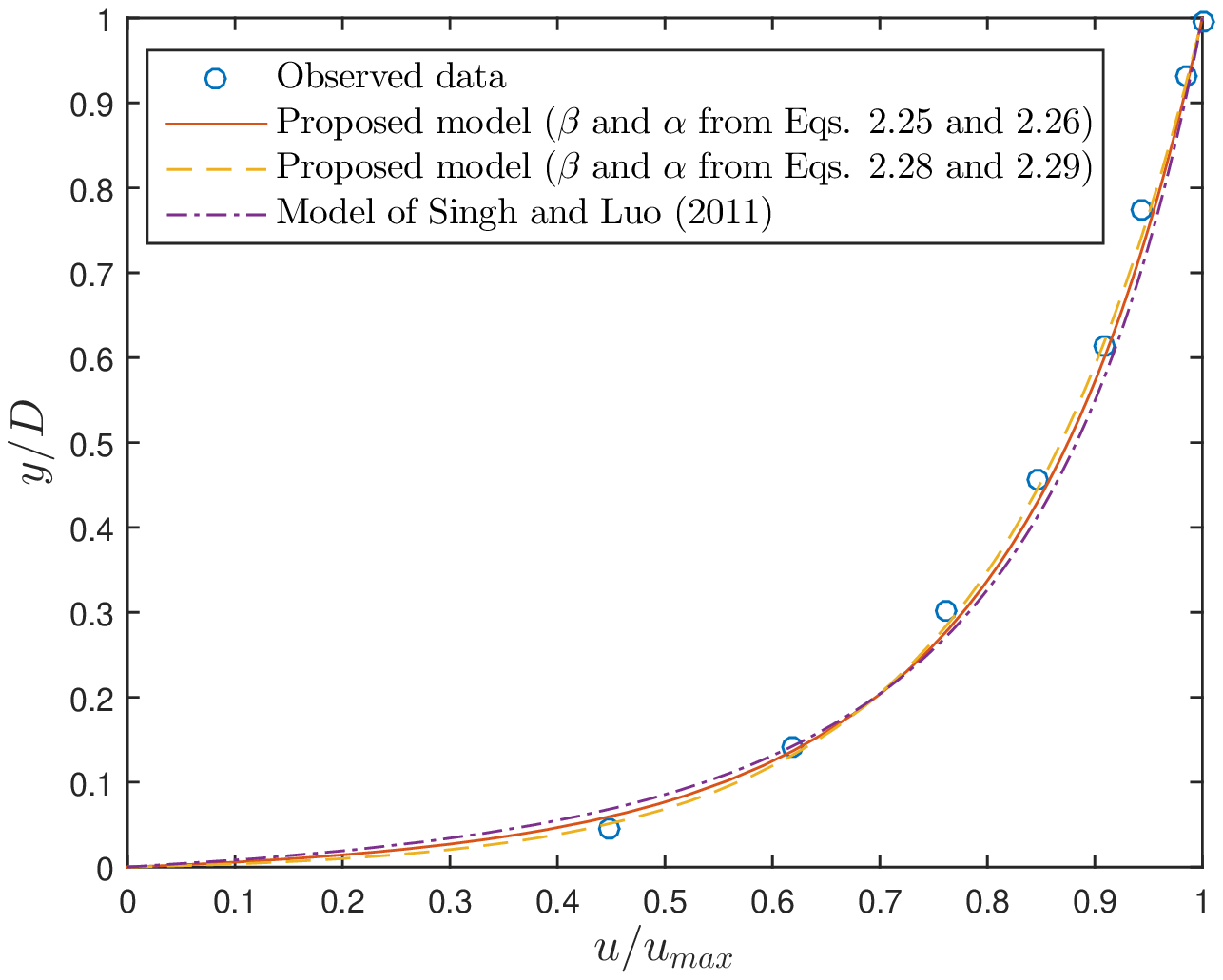}           
                \caption{}
                \label{fig7a}
       \end{subfigure}%
       \hspace{1.5cm}
              \begin{subfigure}[b]{0.21\textwidth}           
                \includegraphics[scale=0.31]{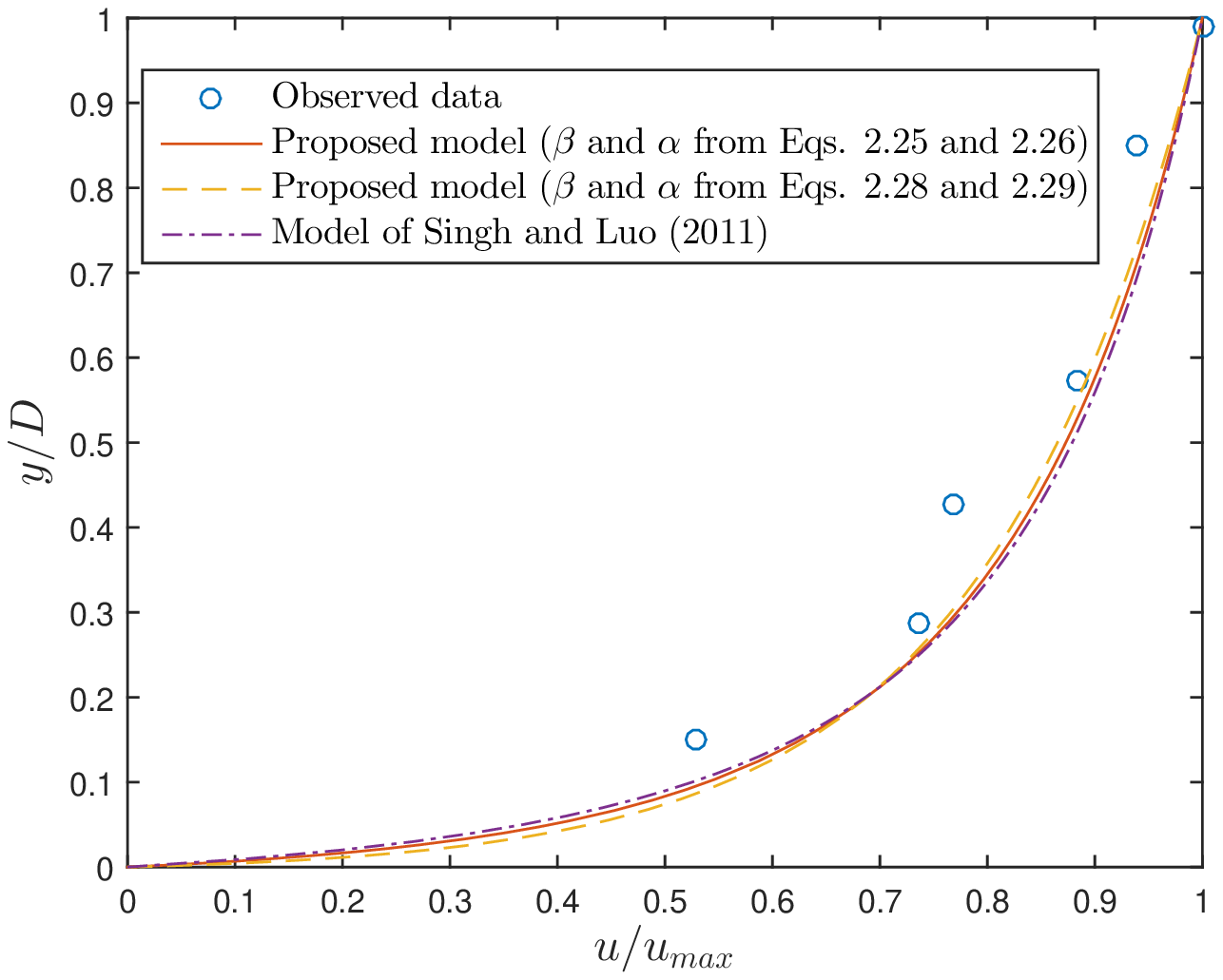}
                \caption{}
                \label{fig7b}
       \end{subfigure}%
 \caption{Comparison of the proposed model with the model of Singh and Luo (\cite{singh2011entropy}) for (a) Run 1, and (b) Run 10 of Davoren (\cite{davoren1985local}) data.}\label{fig7}
\end{figure}

Two sets of data from each of Vanoni (\cite{vanoni1946transportation}), Einstein and Chien (\cite{einstein1955effects}), and Davoren (\cite{davoren1985local}) were selected for the comparison. The numerical solution of the proposed model was considered here for both cases of $\beta$ and $\alpha$ from Eqs. (\ref{eq25})-(\ref{eq26}) and (\ref{eq28})-(\ref{eq29}). The Lagrange multipliers and the entropy index were obtained using the technique mentioned in the previous section, and are reported in Table \ref{t4}. Fig. \ref{fig5} shows the comparison between the models for Run 15 and Run 20 of the Vanoni (\cite{vanoni1946transportation}) data, where it was observed that all three models predicted the velocity values well throughout the water column; however, the proposed model was superior to the model of Singh and Luo (\cite{singh2011entropy}) as can be seen from the figures and RE, RMSE values reported in Table \ref{t6}. In Fig. \ref{fig6}, the models are compared with Run C3 (clear water flow) and Run S5 (sediment-laden flow) of Einstein and Chien (\cite{einstein1955effects}). It is seen from the figure and Table \ref{t6} that the proposed model performed better for Run C3 while for Run S5, the model of Singh and Luo (\cite{singh2011entropy}) was superior. Fig. \ref{fig7} depicts the comparison between the models for field data, namely, Run 1 and Run 10 of Davoren (\cite{davoren1985local}). It can be observed from the figure and Table \ref{t6} that the proposed model predicted the data better than did the model of Singh and Luo (\cite{singh2011entropy}) throughout the water column. Moreover, it can be seen from the figures and the table that the present model with $\beta$ and $\alpha$ from Chow's formulae given by Eqs. (\ref{eq25}) and (\ref{eq26}) outperformed the other models for all the cases except Run S5 data. It may be noted that the proposed model is expected to perform better than it is seen in the present study if one uses a more accurate formula for $\beta$ and $\alpha$, which is not yet available. 

\begin{table}[]
\centering
\caption{Relative error (RE) and the root-mean-squared error (RMSE) of the proposed model and the model and the model of Singh and Luo (\cite{singh2011entropy}) for some selected set of data.}\label{t6}
\scalebox{0.88}{\begin{tabular}{|c|c|c|c|c|c|c|c|}
\hline
\multicolumn{2}{|c|}{\multirow{2}{*}{Data set}} & \multicolumn{2}{c|}{\begin{tabular}[c]{@{}c@{}}$\beta$ and $\alpha$ from\\ Eqs. (\ref{eq25}) and (\ref{eq26})\end{tabular}} & \multicolumn{2}{c|}{\begin{tabular}[c]{@{}c@{}}$\beta$ and $\alpha$ from\\ Eqs. (\ref{eq28}) and (\ref{eq29})\end{tabular}} & \multicolumn{2}{c|}{\begin{tabular}[c]{@{}c@{}}Singh and Luo\\ (\cite{singh2011entropy}) model\end{tabular}} \\ \cline{3-8} 
\multicolumn{2}{|c|}{} & RE & RMSE & RE & RMSE & RE & RMSE \\ \hline
\multirow{2}{*}{Vanoni (\cite{vanoni1946transportation})} & Run 15 & 0.0134 & 0.0159 & 0.0155 & 0.0183 & 0.0569 & 0.1007 \\ \cline{2-8} 
 & Run 20 & 0.0173 & 0.0230 & 0.0248 & 0.0314 & 0.1244 & 0.1888 \\ \hline
\multirow{2}{*}{\begin{tabular}[c]{@{}c@{}}Einstein and \\ Chien (\cite{einstein1955effects})\end{tabular}} & Run C3 & 0.0496 & 0.0534 & 0.0503 & 0.0542 & 0.1796 & 0.2134 \\ \cline{2-8} 
 & Run S5 & 0.1073 & 0.1213 & 0.1213 & 0.1479 & 0.0730 & 0.0949 \\ \hline
\multirow{2}{*}{\begin{tabular}[c]{@{}c@{}}Davoren\\ (\cite{davoren1985local})\end{tabular}} & Run 1 & 0.0161 & 0.0212 & 0.0092 & 0.0138 & 0.0377 & 0.0755 \\ \cline{2-8} 
 & Run 10 & 0.0205 & 0.0292 & 0.0217 & 0.0265 & 0.0626 & 0.0842 \\ \hline
\end{tabular}}
\end{table}

\section{Conclusions}
The following conclusions can be drawn from the present study
\begin{itemize}
  \item The present study derives the vertical distribution of streamwise fluid velocity in wide-open channels using the Tsallis entropy together with the principle of maximum entropy, subject to conservation laws. It also includes the effect of entropy index in the derivation.
  \item The connection between the probability and space domains is made by the variable transformation. The governing differential equation for velocity is solved by homotopy analysis method after converting it to a weaker nonlinear form using Pad{\'e} approximation technique. The convergence of the series solution is tackled by some convergence-control parameters and investigated both theoretically and numerically.
  \item The Lagrange multipliers and the entropy index constitute a system of nonlinear equations after approximating the integrals of constraints using the Gauss-Legendre quadrature rule. The system is then solved by the Gauss-Newton method for assessing the velocity profile.
  \item Laboratory and field data are considered to validate the model and also to compare with the existing velocity profile based on Tsallis entropy. It is found that the incorporation of additional constraints and the effect of entropy index improves the velocity profile. The proposed methodology can be applied further to study different kinds of open channel flow problems.  
\end{itemize}

\appendix
\section{Convergence Analysis of HAM-Based Series Solution Eq. (\ref{eq50})}

\begin{thm}\label{th1}
If the homotopy series $\mathop \sum \limits_{m = 0}^\infty  {\hat u_m}\left( {\hat y} \right)$ and $\mathop \sum \limits_{m = 0}^\infty  \hat u{'_m}\left( {\hat y} \right)$ converge, then ${R_m}\left( {{{\overrightarrow {\hat u} }_{m - 1}}} \right)$ given by Eq. (\ref{eq48}) satisfies the relation $\mathop \sum \limits_{m = 1}^\infty  {R_m}\left( {{{\overrightarrow {\hat u} }_{m - 1}}} \right) = 0$.
\end{thm}
 \textbf{Proof}:
The linear operator is defined as follows
\begin{equation}\label{eqa1}
    \mathcal{L}\left[ {\hat u} \right] = \omega \frac{{d\hat u}}{{d\hat y}}
\end{equation}
According to Eq. (\ref{eq42}), one obtains
\begin{equation}\label{eqa2}
\mathcal{L}\left[ {{{\hat u}_1}} \right] = \hbar {R_1}\left( {{{\overrightarrow {\hat u} }_0}} \right)
\end{equation}
\begin{equation}\label{eqa3}
\mathcal{L}\left[ {{{\hat u}_2} - {{\hat u}_1}} \right] = \hbar {R_2}\left( {{{\overrightarrow {\hat u} }_1}} \right)
\end{equation}
\begin{equation}\label{eqa4}
    \mathcal{L}\left[ {{{\hat u}_3} - {{\hat u}_2}} \right] = \hbar {R_3}\left( {{{\overrightarrow {\hat u} }_2}} \right)
\end{equation}
$$\vdots$$
\begin{equation}\label{eqa5}
\mathcal{L}\left[ {{{\hat u}_m} - {{\hat u}_{m - 1}}} \right] = \hbar {R_m}\left( {{{\overrightarrow {\hat u} }_{m - 1}}} \right)
\end{equation}
Adding all of the above terms, one can get
\begin{equation}\label{eqa6}
\mathcal{L}\left[ {{{\hat u}_m}} \right] = \hbar \mathop \sum \limits_{k = 1}^m {R_k}\left( {{{\overrightarrow {\hat u} }_{k - 1}}} \right)
\end{equation}
As the series $\mathop \sum \limits_{m = 0}^\infty  {\hat u_m}\left( {\hat y} \right)$ and $\mathop \sum \limits_{m = 0}^\infty  \hat u{'_m}\left( {\hat y} \right)$ are convergent, $\lim_{n \to \infty} {\hat u_m}\left( {\hat y} \right) = 0$ and $\lim_{n \to \infty} \hat u{'_m}\left( {\hat y} \right) = 0$. Now, recalling the above summand and taking the limit, the required result follows as
\begin{equation}\label{eqa7}
    \hbar \mathop \sum \limits_{k = 1}^\infty  {R_k}\left( {{{\overrightarrow {\hat u} }_{k - 1}}} \right) =\lim_{m \to \infty} \hbar \mathop \sum \limits_{k = 1}^{m}  {R_k}\left( {{{\overrightarrow {\hat u} }_{k - 1}}} \right) = \lim_{m \to \infty} \mathcal{L} [{\hat{u}_{m}}] = \omega \lim_{m \to \infty} {\hat{u}_{m}}' = 0
\end{equation}

\begin{thm}\label{th2}
If $\hbar$ is so properly chosen that the series $\mathop \sum \limits_{m = 0}^\infty  {\hat u_m}\left( {\hat y} \right)$ and \resizebox{.1\hsize}{!}{$\mathop \sum \limits_{m = 0}^\infty  \hat u{'_m}\left( {\hat y} \right)$} converge absolutely to $\hat u\left( {\hat y} \right)$ and $\hat u'\left( {\hat y} \right)$, respectively, then the homotopy series \resizebox{.1\hsize}{!}{$\mathop \sum \limits_{m = 0}^\infty  {\hat u_m}\left( {\hat y} \right)$} satisfies the original governing Eq. (\ref{eq36}).
\end{thm}
\textbf{Proof}:
Let $\mathop \sum \limits_{i = 0}^\infty  {x_i}$ and $\mathop \sum \limits_{j = 0}^\infty  {y_j}$ be two infinite series of real/complex terms. Then the Cauchy product of the above two series is defined by the discrete convolution as follows
\begin{equation}\label{eqa8}
    \left( {\mathop \sum \limits_{i = 0}^\infty  {x_i}{\rm{\;}}} \right)\left( {\mathop \sum \limits_{j = 0}^\infty  {y_j}} \right) = \mathop \sum \limits_{k = 0}^\infty  \mathop \sum \limits_{l = 0}^k {x_l}{y_{k - l}}
\end{equation}
Therefore, using the above rule in relation to Eq. (\ref{eq48}), we get
\begin{equation}\label{eqa9}
\mathop \sum \limits_{m = 1}^\infty  \mathop \sum \limits_{j = 0}^{m - 1} {\hat u_j}{\hat u_{m - 1 - j}} = {\left( {\mathop \sum \limits_{m = 0}^\infty  {{\hat u}_m}{\rm{\;}}} \right)^2}
\end{equation}
\begin{equation}\label{eqa10}
\mathop \sum \limits_{m = 1}^\infty  \mathop \sum \limits_{j = 0}^{m - 1} {\hat u_j}\hat u{'_{m - 1 - j}} = \left( {\mathop \sum \limits_{m = 0}^\infty  {{\hat u}_m}{\rm{\;}}} \right)\left( {\mathop \sum \limits_{k = 0}^\infty  \hat u{'_k}} \right)
\end{equation}
\begin{equation}\label{eqa11}
\mathop \sum \limits_{m = 1}^\infty  \mathop \sum \limits_{j = 0}^{m - 1} {\hat u_{m - 1 - j}}\mathop \sum \limits_{k = 0}^j {\hat u_k}{\hat u_{j - k}} = {\left( {\mathop \sum \limits_{m = 0}^\infty  {{\hat u}_m}{\rm{\;}}} \right)^3}
\end{equation}
\begin{equation}\label{eqa12}
\mathop \sum \limits_{m = 1}^\infty  \mathop \sum \limits_{j = 0}^{m - 1} {\hat u_{m - 1 - j}}\mathop \sum \limits_{k = 0}^j {\hat u_k}\hat u{'_{j - k}} = {\left( {\mathop \sum \limits_{m = 0}^\infty  {{\hat u}_m}{\rm{\;}}} \right)^2}\left( {\mathop \sum \limits_{k = 0}^\infty  \hat u{'_k}} \right)
\end{equation}
\begin{equation}\label{eqa13}
\mathop \sum \limits_{m = 1}^\infty  \mathop \sum \limits_{j = 0}^{m - 1} {\hat u_{m - 1 - j}}\mathop \sum \limits_{k = 0}^j {\hat u_{j - k}}\mathop \sum \limits_{l = 0}^k {\hat u_l}{\hat u_{k - l}} = {\left( {\mathop \sum \limits_{m = 0}^\infty  {{\hat u}_m}{\rm{\;}}} \right)^4}
\end{equation}
\begin{equation}\label{eqa14}
\mathop \sum \limits_{m = 1}^\infty  \mathop \sum \limits_{j = 0}^{m - 1} {\hat u_{m - 1 - j}}\mathop \sum \limits_{k = 0}^j {\hat u_{j - k}}\mathop \sum \limits_{l = 0}^k {\hat u_l}\hat u{'_{k - l}} = {\left( {\mathop \sum \limits_{m = 0}^\infty  {{\hat u}_m}{\rm{\;}}} \right)^3}\left( {\mathop \sum \limits_{k = 0}^\infty  \hat u{'_k}} \right)
\end{equation}
\begin{equation}\label{eqa15}
\mathop \sum \limits_{m = 1}^\infty  \mathop \sum \limits_{j = 0}^{m - 1} {\hat u_{m - 1 - j}}\mathop \sum \limits_{k = 0}^j {\hat u_{j - k}}\mathop \sum \limits_{l = 0}^n {\hat u_{n - l}}\mathop \sum \limits_{p = 0}^l {\hat u_p}\hat u{'_{l - p}} = {\left( {\mathop \sum \limits_{m = 0}^\infty  {{\hat u}_m}{\rm{\;}}} \right)^4}\left( {\mathop \sum \limits_{k = 0}^\infty  \hat u{'_k}} \right)
\end{equation}
Theorem \ref{th1} shows that if $\mathop \sum \limits_{m = 0}^\infty  {\hat u_m}\left( {\hat y} \right)$ and $\mathop \sum \limits_{m = 0}^\infty  \hat u{'_m}\left( {\hat y} \right)$ converge then \resizebox{.2\hsize}{!}{$\mathop \sum \limits_{m = 1}^\infty  {R_m}\left( {{{\overrightarrow {\hat u} }_{m - 1}}} \right) = 0$}. Therefore, substituting the above expressions in Eq. (\ref{eq48}) and simplifying further lead to
\begin{align}\label{eqa16}
    & \mathop \sum \limits_{m = 0}^\infty  \hat u{'_m} + {D_1}\left( {\mathop \sum \limits_{m = 0}^\infty  \hat u_m^{'}} \right)\left( {\mathop \sum \limits_{k = 0}^\infty  {{\hat u}_k}{\rm{\;}}} \right) + {D_2}\left( {\mathop \sum \limits_{m = 0}^\infty  \hat u_m^{'}} \right){\left( {\mathop \sum \limits_{k = 0}^\infty  {{\hat u}_k}{\rm{\;}}} \right)^2} \nonumber \\& + {D_3}\left( {\mathop \sum \limits_{m = 0}^\infty  \hat u_m^{'}} \right){\left( {\mathop \sum \limits_{k = 0}^\infty  {{\hat u}_k}{\rm{\;}}} \right)^3} + {D_4}\left( {\mathop \sum \limits_{m = 0}^\infty  \hat u_m^{'}} \right){\left( {\mathop \sum \limits_{k = 0}^\infty  {{\hat u}_k}{\rm{\;}}} \right)^4} - {C_0}\mathop \sum \limits_{m = 0}^\infty  \left( {1 - {\chi _{m + 1}}} \right) \nonumber \\& - {C_1}\mathop \sum \limits_{m = 0}^\infty  {\hat u_m} - {C_2}{\left( {\mathop \sum \limits_{k = 0}^\infty  {{\hat u}_k}{\rm{\;}}} \right)^2} - {C_3}{\left( {\mathop \sum \limits_{k = 0}^\infty  {{\hat u}_k}{\rm{\;}}} \right)^3} - {C_4}{\left( {\mathop \sum \limits_{k = 0}^\infty  {{\hat u}_k}{\rm{\;}}} \right)^4} = 0
\end{align}
which is basically the original governing equation Eq. (\ref{eq36}). Furthermore, subject to the initial condition ${\hat u_0}\left( 0 \right) = 0$ and the conditions for the higher-order deformation equation ${\hat u_m}\left( 0 \right) = 0$, for $m \ge 1$, we easily obtain $\mathop \sum \limits_{m = 0}^\infty  {\hat u_m}\left( 0 \right) = 0$. Hence, the convergence result follows.

\subsection*{Acknowledgment}
The first two authors are thankful to the Science and Engineering Research Board (SERB), Department of Science and Technology (DST), Govt. of India for providing the financial support through the research project with no.: SERB/F/4873/2018-2019, Dated 24 July 2018.

\thispagestyle{empty}

\bibliographystyle{unsrt}

\begin{thebibliography}{1}
\bibitem{chow1959open} Ven T. Chow, \textit{Open-channel hydraulics}, McGrawHill, 1959.
\bibitem{vanoni2006sedimentation} Vito A. Vanoni, \textit{Sedimentation engineering}, American Society of Civil Engineers, 2006.
\bibitem{prandtl19257} L. Prandtl, \textit{bericht uber untersuchungen zur ausgebildeten turbulenz} Zeitschrift fur Angewandte Mathematik und Mechanik, \textbf{5(2)} (1925), 136–-139.
\bibitem{nezu2005open} I. Nezu, \textit{Open-channel flow turbulence and its research prospect in the 21st century} Journal of Hydraulic Engineering, \textbf{131(4)} (2005), 229-–246.
\bibitem{coles1956law} D. Coles, \textit{The law of the wake in the turbulent boundary layer} Journal of Fluid Mechanics, \textbf{1(2)} (1956), 191–-226.
\bibitem{guo2003modified} J. Guo and P. Y. Julien, \textit{Modified log-wake law for turbulent flow in smooth pipes} Journal of Hydraulic Research \textbf{41(5)} (2003), 493–-501.
\bibitem{yang2005investigation} S. Yang, S. Lim, and J.A. McCorquodale, \textit{Investigation of near wall velocity in 3-d smooth channel flows} \textit{Journal of Hydraulic Research} \textbf{43(2)} (2005), 149–-157.
\bibitem{afzal2005scaling} N. Afzal, \textit{Scaling of power law velocity profile in wall-bounded turbulent shear flows} 43rd AIAA Aerospace Sciences Meeting and Exhibit, page 109, 2005.
\bibitem{yang2004velocity} S. Yang, S. Tan, and S. Lim, \textit{Velocity distribution and dip-phenomenon in smooth uniform open channel flows} Journal of Hydraulic Engineering, \textbf{130(12)} (2004), 1179-–1186.
\bibitem{kundu2012analytical} S. Kundu and K. Ghoshal, \textit{An analytical model for velocity distribution and dip-phenomenon in uniform open channel flows} International Journal of Fluid Mechanics Research, \textbf{39(5)} (2012).
\bibitem{chiu1987entropy} C. L. Chiu, \textit{Entropy and probability concepts in hydraulics}. Journal of Hydraulic Engineering, \textbf{113(5)} (1987) 583–-599.
\bibitem{shannon1948mathematical}  Claude E. Shannon, \textit{A mathematical theory of communication}. Bell system technical journal, \textbf{27(3)} (1948) 379-–423.
\bibitem{jaynesa1957information} E. T. Jaynes, \textit{Information theory and statistical mechanics} Physical Review \textbf{106(4)} (1957) 620.
\bibitem{jaynesb1957information} E. T. Jaynes, \textit{Information theory and statistical mechanics. II.} Physical Review \textbf{108(2)} (1957) 171.
\bibitem{kapur1989maximum} J. N. Kapur, \textit{Maximum-entropy models in science and engineering} John Wiley \& Sons, 1989.
\bibitem{chiu1995maximum} C.L. Chiu and C. A. A. Said, \textit{Maximum and mean velocities and entropy in open-channel flow}. Journal of Hydraulic Engineering, \textbf{121(1)} (1995) 26-–35.
\bibitem{cao1997entropy} S. Cao and D. W. Knight, \textit{Entropy-based design approach of threshold alluvial channels} Journal of Hydraulic Research, \textbf{35(4)} (1997) 505-–524.
\bibitem{chiu2000mathematical} C.L. Chiu, W. Jin, and Y.C. Chen. \textit{Mathematical models of distribution of sediment concentration} Journal of Hydraulic Engineering, \textbf{126(1)} (2000) 16–-23.
\bibitem{sterling2002attempt} M. Sterling and D. Knight, \textit{An attempt at using the entropy approach to predict the transverse distribution of boundary shear stress in open channel flow.} Stochastic Environmental Research and Risk Assessment, \textbf{16(2)} (2002) 127-–142.
\bibitem{araujo2007entropy}  J. C. de Araujo, \textit{Entropy-based equation to assess hillslope sediment production}. Earth Surface Processes and Landforms, \textbf{32(13)} (2007) 2005-–2018.
\bibitem{moramarco2010formulation} T. Moramarco and V. P. Singh, \textit{Formulation of the entropy parameter based on hydraulic and geometric characteristics of river cross sections} Journal of Hydrologic Engineering, \textbf{15(10)} (2010) 852-–858.
\bibitem{bechle2014entropy} A. J. Bechle and C. H. Wu, \textit{An entropy-based surface velocity method for estuarine discharge measurement}. Water Resources Research, \textbf{50(7)} (2014) 6106-–6128.
\bibitem{greco2014entropy} M. Greco and D. Mirauda, \textit{Entropy parameter estimation in large-scale roughness open channel}. Journal of Hydrologic Engineering, \textbf{20(2)} (2014) 04014047.
\bibitem{kundu2017prediction} S. Kundu, \textit{Prediction of velocity-dip-position over entire cross section
of open channel flows using entropy theory}. Environmental Earth Sciences,
\textbf{76(10)}, (2017), 363.
\bibitem{singh2014entropy} Vijay P. Singh, \textit{Entropy theory in hydraulic engineering: an introduction}. American Society of Civil Engineers, 2014.
\bibitem{tsallis1988possible} Constantino Tsallis, \textit{Possible generalization of boltzmann-gibbs statistics}. Journal of statistical physics, \textbf{52(1-2)}, (1988), 479-–487.
\bibitem{singh2011entropy} Vijay P. Singh and H. Luo, \textit{Entropy theory for distribution of one-dimensional velocity in open channels}. Journal of Hydrologic Engineering, \textbf{16(9)}, (2011), 725-–735.
\bibitem{luo2010entropy} H. Luo and Vijay P. Singh, \textit{Entropy theory for two-dimensional velocity distribution}. Journal of Hydrologic Engineering, \textbf{16(4)}, (2010), 303-–315.
\bibitem{cui2013one} H. Cui and Vijay P. Singh, \textit{One-dimensional velocity distribution in open channels using tsallis entropy}. Journal of Hydrologic Engineering, \textbf{19(2)}, (2013), 290–-298.
\bibitem{cui2012two}H. Cui and Vijay P. Singh, \textit{Two-dimensional velocity distribution in open channels using the tsallis entropy}. Journal of Hydrologic Engineering, \textbf{18(3)} (2012), 331–-339.
\bibitem{koutsoyiannisa2005uncertainty} D. Koutsoyiannis, \textit{Uncertainty, entropy, scaling and hydrological stochastics. 1. marginal distributional properties of hydrological processes and state scaling}. Hydrological Sciences Journal, \textbf{50(3)}, 2005.
\bibitem{koutsoyiannisb2005uncertainty} D. Koutsoyiannis, \textit{Uncertainty, entropy, scaling and hydrological stochastics. 2. time dependence of hydrological processes and time scaling}. Hydrological Sciences Journal, \textbf{50(3)}, 2005.
\bibitem{keylock2005describing} C.J. Keylock, \textit{Describing the recurrence interval of extreme floods using nonextensive thermodynamics and tsallis statistics}. Advances in Water Resources, \textbf{28(8)} (2005), 773-–778.
\bibitem{bonakdari2015comparison} H. Bonakdari, Z. Sheikh, and M. Tooshmalani, \textit{Comparison between shannon and tsallis entropies for prediction of shear stress distribution in open channels}. Stochastic Environmental Research and Risk Assessment, \textbf{29(1)} (2015), 1-–11.
\bibitem{singh2016introduction} Vijay P Singh, \textit{Introduction to tsallis entropy theory in water engineering}. CRC Press, 2016.
\bibitem{gholami2019method} A. Gholami, H. Bonakdari, and A. Mohammadian, \textit{A method based on the tsallis entropy for characterizing threshold channel bank profiles}. Physica A: Statistical Mechanics and its Applications, \textbf{526}, 2019.
\bibitem{tsallis1999nonextensive} C. Tsallis, \textit{Nonextensive statistics: theoretical, experimental and computational evidences and connections}. Brazilian Journal of Physics, \textbf{29(1)} (1999), 1–-35.
\bibitem{lyra1998nonextensivity} M.L. Lyra and C. Tsallis, \textit{Nonextensivity and multifractality in low-dimensional dissipative systems}. Physical Review Letters, \textbf{80(1)} (1998), 53.
\bibitem{conroy2015determining} J.M. Conroy and H.G. Miller, \textit{Determining the tsallis parameter via maximum entropy}. Physical Review E, \textbf{91(5)} (2015), 052112.
\bibitem{plastino1999tsallis} A.R.P.A. Plastino and A.R. Plastino, \textit{Tsallis entropy and jaynes’ information theory formalism}. Brazilian Journal of Physics, \textbf{29(1)} (1999), 50–-60.
\bibitem{chiu1989velocity} C.L. Chiu, \textit{Velocity distribution in open channel flow}. Journal of Hydraulic Engineering, \textbf{115(5)} (1989), 576-–594.
\bibitem{kumbhakar2019application} M. Kumbhakar, K. Ghoshal, and Vijay P. Singh, \textit{Application of relative entropy theory to streamwise velocity profile in open-channel flow: effect of prior probability distributions}. Zeitschrift fur angewandte Mathematik und Physik, \textbf{70(3)} (2019), 80.
\bibitem{kelley2003solving} C. T. Kelley, \textit{Solving nonlinear equations with Newton’s method}, Volume 1, Siam, 2003.
\bibitem{martinez2000practical} J. M. Martınez, \textit{Practical quasi-newton methods for solving nonlinear systems}. Journal of Computational and Applied Mathematics, \textbf{124(1-2)} (2000), 97-–121.
\bibitem{nocedal2006numerical} J. Nocedal and S. Wright, \textit{Numerical optimization}. Springer Science \& Business Media, 2006.
\bibitem{gratton2007approximate} S. Gratton, A. S. Lawless, and Nancy K. Nichols, \textit{Approximate gauss–newton methods for nonlinear least squares problems}. SIAM Journal on Optimization, \textbf{18(1)} (2007), 106-–132.
\bibitem{chiu2006probabilistic} C. L. Chiu and Shih-Meng Hsu, \textit{Probabilistic approach to modeling of velocity distributions in fluid flows}. Journal of Hydrology, \textbf{316(1-4)} (2006), 28-–42.
\bibitem{baker1996pade} G. A. Baker and Peter Graves Morris, \textit{Pad{\'e} approximants}, volume 59. Cambridge University Press, 1996.
\bibitem{liao1992proposed} Shijun Liao, \textit{The proposed homotopy analysis technique for the solution of nonlinear problems}. PhD thesis, Shanghai Jiao Tong University Shanghai, 1992.
\bibitem{liao2012homotopy} Shijun Liao, \textit{Homotopy analysis method in nonlinear differential equations} Springer, 2012.
\bibitem{vajravelu2013nonlinear} K. Vajravelu and R. Van Gorder, \textit{Nonlinear flow phenomena and homotopy analysis} Springer, 2013.
\bibitem{nave2013comparison} O. Nave, S. Ajadi, Y. Lehavi, and V. Goldshtein, \textit{Comparison of the homotopy perturbation method (hpm) and method of integral manifolds (mim) on a thermal explosion of polydisperse fuel spray system}. SIAM Journal on Applied Mathematics, \textbf{73(2)} (2013), 929-–952.
\bibitem{vanoni1946transportation} Vito A Vanoni, \textit{Transportation of suspended sediment by water}. Trans. of ASCE, \textbf{111} (1946), 67-–102.
\bibitem{einstein1955effects} H.A. Einstein and N. Chien, \textit{Effects of heavy sediment concentration near the bed on velocity and sediment distribution}, mrd sediment ser. 8. Univ. of Calif., Berkeley, 1955.
\bibitem{davoren1985local} A. Davoren, \textit{Local scour around a cylindrical bridge pier}. Hydrology Centre, Ministry of Works and Development for the National Water, 1985.
\end{thebibliography}

\end{document}